\documentclass[12pt]{article}
\pdfoutput=1

\usepackage[english]{babel}
\usepackage[latin1]{inputenc}
\usepackage[intlimits]{amsmath}
\usepackage{amssymb}
\usepackage{url}
\usepackage{graphicx}
\usepackage{slashed}
\usepackage{color}
\usepackage{booktabs}
\usepackage[small]{caption}
\usepackage{cite}

\numberwithin{equation}{section}

\DeclareMathOperator{\BR}{BR}

\newcommand{\GeV}{\mbox{GeV}}

\newcommand{\cm}{\mbox{cm}}
\newcommand{\Mp}{M_{\mathrm{P}}}

\begin{document}

\begin{flushleft}
DESY 12-112\\
June 2012
\end{flushleft}

\vskip 1cm

\begin{center}
{\LARGE\bf Decaying vs Annihilating Dark Matter\\
in Light of a Tentative Gamma-Ray Line\\ [2mm]}

\vskip 2cm

{\large Wilfried~Buchm\"uller and Mathias~Garny}\\[3mm]
{\it{
Deutsches Elektronen-Synchrotron DESY, 22607 Hamburg, Germany}
}
\end{center}

\vskip 1cm

\begin{abstract}
\noindent
Recently reported tentative evidence for a gamma-ray line in the Fermi-LAT
data is of great potential interest for identifying the nature of dark matter.
We compare the implications for decaying and annihilating dark matter 
taking the constraints from continuum gamma-rays, antiproton
flux and morphology of the excess into account. We find that higgsino
and wino dark matter are excluded, also for nonthermal production.
Generically, the continuum gamma-ray flux severely constrains annihilating
dark matter. Consistency of decaying dark matter with the
spatial distribution of the Fermi-LAT excess would require an enhancement
of the dark matter density near the Galactic center.
\end{abstract}

\thispagestyle{empty}

\newpage

\section{Introduction}

Monochromatic gamma-ray lines have been suggested long ago as signature for
pair-annihilation of dark matter (DM) particles \cite{Bergstrom:1988fp}.
During the past decades many DM candidates have been discussed, the most 
popular ones being weakly interacting massive particles (WIMPs) in the context
of supersymmetric extensions of the Standard Model \cite{Bertone:2004pz}. More
recently, also decaying dark matter has been studied in detail 
\cite{Takayama:2000uz,Buchmuller:2007ui,Cirelli:2012ut,Arvanitaki:2008hq}.
An attractive feature of decaying gravitino dark matter is the consistency 
with thermal leptogenesis, contrary to standard WIMP dark matter 
\cite{Buchmuller:2007ui}.

During the past years the Large Area Telescope (LAT)~\cite{Atwood:2009ez},
on board the Fermi gamma-ray space telescope, has searched with 
unprecedented sensitivity for photon lines from 30~MeV to 300~GeV. 
Stringent constraints on decaying and annihilating dark matter have been
obtained from searches in the energy range $30-200$~GeV based on 11 months of 
data~\cite{Abdo:2010nc}, and in the range $7-200$~GeV based on 23 months of 
data~\cite{Ackermann:2012qk}. The search region used in these analyses 
covers the whole sky except for the Galactic disk ($|b|>10^\circ$) plus a 
$20^\circ\times 20^\circ$ region around the Galactic center. A similar analysis
has been performed independently based on publicly available data corresponding
to 27 months in Ref.~\cite{Vertongen:2011mu}. No indications for gamma-ray lines 
were found. 

In a recent analysis, that is based on optimized search regions around the
Galactic center, and takes 43 months of Fermi-LAT data into account, a hint
for a gamma-ray feature in the energy range $120-140$~GeV is reported~\cite
{Bringmann:2012vr,Weniger:2012tx}. When interpreted in terms of a gamma-ray
line~\cite{Weniger:2012tx}, the significance obtained from the statistical
uncertainties in the search regions close to the Galactic center is $4.6\sigma$.
The significance is reduced to $3.3\sigma$ when correcting for the bias
introduced by selecting the search regions. While the excess is currently under
an active debate~\cite{Profumo:2012tr,Tempel:2012ey,Boyarsky:2012ca,
Bergstrom:2012fi,Ibarra:2012dw,Dudas:2012pb}, the claim has been further
strengthened by a recent analysis~\cite{Su:2012ft}, which confirms the existence
of an excess, as well as its spectral shape, with even higher statistical
significance than the one claimed in Ref.~\cite{Weniger:2012tx}. In addition,
indications are found that the excess originates from a relatively narrow region
of a few degrees around the Galactic center, possibly with a small offset within
the Galactic plane. Hopefully, the question about the existence of the spectral
feature and its precise properties will be settled in the near future. 

Assuming that the feature is real, the question about its origin is of great
interest. While an astrophysical explanation might eventually be identified,
a spectral feature in this energy range can arise rather generically from the
annihilation or decay of dark matter particles. In this paper we compare decaying
and annihilating dark matter based on two prototype models: decaying gravitinos
and wino/higgsino-like WIMPs which annihilate predominantly into two W-bosons. In
both cases we treat the branching ratios into $\gamma\nu$ (gravitino) and
$\gamma\gamma$ (wino/higgsino) final states as free parameters to account for
some model dependence. Note that we do not demand thermal freeze-out for WIMPs. 
Higgsinos and winos can be nonthermally produced in gravitino decays, compatible
with leptogenesis \cite{Ibe:2011aa,Buchmuller:2012bt}, or, alternatively, in
moduli decays~\cite{Moroi:1999zb,Kitano:2008tk,Acharya:2012tw}.

In the following we shall compare interpretations of the tentative 130~GeV
photon line in terms of decaying and annihilating dark matter in a sequence
of increasing assumptions: We first consider constraints from continuum
gamma-rays, which are independent of charged cosmic rays and the dark matter
distribution (Section~2). This is followed by a discussion of constraints 
from antiprotons, which depend on the propagation model (Section~3). We
then analyze the implications of the spatial distribution of the Fermi-LAT 
excess (Section~4) and draw our conclusions (Section~5).

\section{Constraints from continuum gamma-rays}

Since the dark matter particle is required to be electrically neutral, the
annihilation or decay into photons is typically suppressed compared to channels
involving for example electroweak gauge bosons or the Higgs boson, as well as
quarks or leptons. The subsequent decay and fragmentation of these annihilation
or decay products gives rise, among others, to an emission of gamma rays with a
broad spectrum in the energy range $\simeq 1-100$ GeV below the line. Therefore,
one expects that the line flux is accompanied by an associated continuum flux
with the same spatial distribution, whose strength depends on the relative size
of the annihilation or decay rates into photons and into other Standard Model
particles. As compared to the sharp feature resulting from the direct decay into
photons, the continuum emission has a less characteristic spectral shape and is
therefore much harder to disentangle from the background gamma-ray flux in the
Galactic center region~\cite{Boyarsky:2010dr}. On the other hand, one expects a
much larger total flux in the continuum compared to the line, and therefore it is
important to check up to which level a continuum emission is acceptable. Apart
from the center of the Galaxy, the continuum flux of gamma rays can be
constrained, {\it e.g.}, by the emission seen within the Galactic
halo~\cite{Ackermann:2012rg}, from Galaxy
clusters~\cite{Ackermann:2010rg,Ripken:2010ja}, or from dwarf
galaxies~\cite{Ackermann:2011wa}. In general, a comparison of the various
constraints from different targets is affected by the uncertainties related to the
dark matter density distribution in the different environments.

We will not enter into the details of this discussion here, but instead follow a
very conservative approach. Namely, we  consider the same data sets for the
photon flux that have been used in Ref.~\cite{Weniger:2012tx} to search for a
gamma-ray line signal. However, instead of using only a line as the template,
we add a contribution from the continuum flux, and test whether the combined
spectrum yields a satisfactory fit to the data. This strategy has the advantage
that it is not affected by uncertainties related to the actual dark matter
distribution, because both the line and the continuum fluxes are generated
from the same source. Concretely, we consider a spectral template for the fit
given by
\begin{equation}\label{eq:templateDecayingDM}
\frac{dJ}{dE} = \alpha \left( \delta(E-E_\gamma) + \frac{dN_{EG}}{dE}  + \frac{1-\BR_{\gamma}}{N_\gamma\BR_{\gamma}} \frac{dN^\gamma_{cont}}{dE} \right) + \beta \left(\frac{E}{E_\gamma}\right)^{-\gamma} \,,
\end{equation}
where $E_\gamma=m_{DM}(m_{DM}/2)$,
$\BR_\gamma=\sigma v_{\gamma\gamma}/\sigma v_{tot}\,(\tau/\tau_{\gamma\nu})$
and $N_\gamma=2(1)$ for dark matter annihilation (decay). The spectrum of
secondary photons originating from the decay and fragmentation of quarks,
leptons or electroweak gauge bosons produced in annihilation or decay
processes is denoted by $dN^\gamma_{cont}/dE$, while $dN_{EG}/dE$ denotes
the extragalactic contribution resulting from the superposition of
redshifted photons~\cite{Bertone:2007aw}. Both will be discussed in detail
below.
The fit parameter $\alpha$ determines the strength of the dark matter flux.
The background is assumed to follow a power law within the considered energy
range (see~\cite{Profumo:2012tr} for possible caveats). Both the normalization
$\beta$ and the slope $\gamma$ of the background $(\propto \beta E^{-\gamma})$
are taken as fit parameters.
For a given dark matter profile and search region, $\alpha$ can be
easily converted into the dark matter annihilation cross section
$\sigma v_{\gamma\gamma}$ or partial lifetime $\tau_{\gamma\nu}$ into
monochromatic photons. The continuum spectrum is in general given by the sum of
two components. The first one arises from redshifted monochromatic photons with
extragalactic origin. For decaying dark matter it is given
by~\cite{Bertone:2007aw}
\begin{equation}\label{eq:lineEG}
\frac{dN_{EG}}{dE} = \frac{\Omega_{DM}\rho_{c0}}{\sqrt{\Omega_M}(H_0/c)\bar J_\psi} \, \frac{E^{1/2}}{E_\gamma^{3/2}}\left(1+\frac{\Omega_\Lambda}{\Omega_M}\left(\frac{E}{E_\gamma}\right)^3\right)^{-1/2} \Theta(E_\gamma-E)\;.
\end{equation}
Here $\Omega_\Lambda=1-\Omega_M\simeq 0.73$ are the cosmological density parameters,
assuming a spatially flat $\Lambda$CDM model, $\Omega_{DM}h^2=0.112$ is the dark
matter density, $H_0=h 100$km/s/Mpc is the Hubble constant with $h=0.704$,
$\rho_{c0}=1.05\cdot 10^{-5} h^2$GeV/cm$^3$ is the critical density, $c$ is the
speed of light, and the factor
$\bar J_\psi=\Delta\Omega^{-1}\int_{\Delta\Omega}d\Omega\int_{l.o.s.}\rho_{dm}(r)$
appears because of the normalization chosen in (\ref{eq:templateDecayingDM}).
For the search regions 3 and 4 of~\cite{Weniger:2012tx} and the various density
profiles, the prefactor in (\ref{eq:lineEG}) is of the order $0.3-0.4$. 

The second contribution to the continuum gamma spectrum arises from the weighted
sum of the spectra produced by annihilation or decays, excluding
$\gamma\gamma(\gamma\nu)$,
\begin{equation}
\frac{dN^\gamma_{cont}}{dE} \equiv \frac{1}{\sum_{f\not=\gamma} \BR_f} \sum_{f\not=\gamma} \BR_f \frac{ dN^\gamma_f}{dE} \;,
\end{equation}
where $dN^\gamma_f/dE$ is the number of photons per energy and per
annihilation/decay resulting from each mode, and $\BR_f$ is the corresponding
branching ratio. For the line, we use the same shape as in~\cite{Weniger:2012tx}
based on the instrument response function of the Fermi-LAT detector
(FWHM$=0.136$), which has somewhat broader tails compared to a Gaussian, and
also smooth the step function appearing in the extragalactic contribution
(\ref{eq:lineEG}) accordingly. The continuum flux is determined using
PYTHIA 8.1~\cite{Sjostrand:2007gs}. In order to increase the sensitivity to the
continuum flux we use the data presented in~\cite{Weniger:2012tx} over the full
energy range $20-300$ GeV, taking the more finely binned data where they are
available ($80-200$ GeV). We show our results for both the SOURCE and ULTRA\-CLEAN
event classes as defined in the Pass 7 Version 6 data release, corresponding to
43 months of data~\cite{Weniger:2012tx}. The former class yields better statistics
due to an increased effective area, and the latter a lower contamination with
charged cosmic rays. For both, the zenith-angle cut is $\theta<100^\circ$.
As discussed above, the constraints arising from the relative strength of
monochromatic and continuum gamma rays are rather insensitive to the dark
matter profile. For definiteness, we assume an Einasto
profile~\cite{Navarro:2003ew} normalized to a local density
$\rho_0=0.4\GeV/\cm^3$ unless stated otherwise (see Section~\ref{sec:Profile}
for details). Lastly, we require that $\gamma>2.0$ as a conservative
lower limit for the background power law index \cite{Su:2010qj}, while its
normalization is freely varied. For performing the fit we use the profile
likelihood method as detailed in~\cite{Weniger:2012tx}, with $95\%$C.L.
statistical errors obtained by varying $\alpha$ (with fixed $\BR_\gamma$) and
profiling over $\beta$ and $\gamma$ until the $TS$-value has decreased by 3.84.
We checked that our statistical analysis reproduces the best-fit values and
confidence intervals in~\cite{Weniger:2012tx} using the smaller energy window
and a single line up to $10\%$ for regions 2-4, both data samples and all
density profiles. When including $\BR_\gamma$ in the fit, we use
$\Delta(TS)=5.99$ to obtain the $95\%$C.L. regions, corresponding to the
$\chi^2_{k=2}$ distribution that is expected for two parameters of
interest\footnote{By generating a large sample of data that would be expected
in the presence of a true signal, we checked that the actual distribution of
the profile likelihood ratio is well described by $\chi^2_{k=1}$ when keeping
$\BR_\gamma$ fixed, while it is somewhat steeper than $\chi^2_{k=2}$ when
including $\BR_\gamma$ in the fit, lying between a $\chi^2_{k=1}$ and
$\chi^2_{k=2}$ distribution. Therefore the confidence regions assuming
$\chi^2_{k=2}$ may be regarded as conservative.}. We will comment on the
dependence on the various assumptions later on.

\subsection{Decaying dark matter}

For decaying gravitino dark matter, which we consider as representative example,
we take the decay channels $Z\nu,W\ell,h\nu$ apart from $\gamma\nu$ into account.
In the appendix, the partial decay widths are given in the case of bilinear R-parity
breaking as functions of the gravitino mass, parameters of R-parity breaking, gaugino
masses and the higgsino mass parameter $\mu$. According to Eq.~(\ref{line}), the 
branching ratio into $\gamma\nu$ is enhanced for hierarchical gaugino masses. 
For example, in the case of wino NLSP with $M_2 \simeq m_{3/2}$ one obtains the
maximal branching ratio,
\begin{equation}\label{BRmax}
\BR^{\rm max}_\gamma \simeq \frac{3\pi\alpha}{2\sqrt{2}G_Fm_{3/2}^2} \ ,
\end{equation}
which yields $\BR^{\rm max}_\gamma \simeq 3\%$ for $m_{3/2} \simeq 260$~GeV.
On the contrary, for higgsino NLSP, with $\mu \simeq m_{3/2} \ll M_1,M_2$,
the branching ratio into $\gamma\nu$ is negligible.

\begin{figure}
\vspace*{-1.5cm}
\begin{center}
 \includegraphics[width=0.9\textwidth]{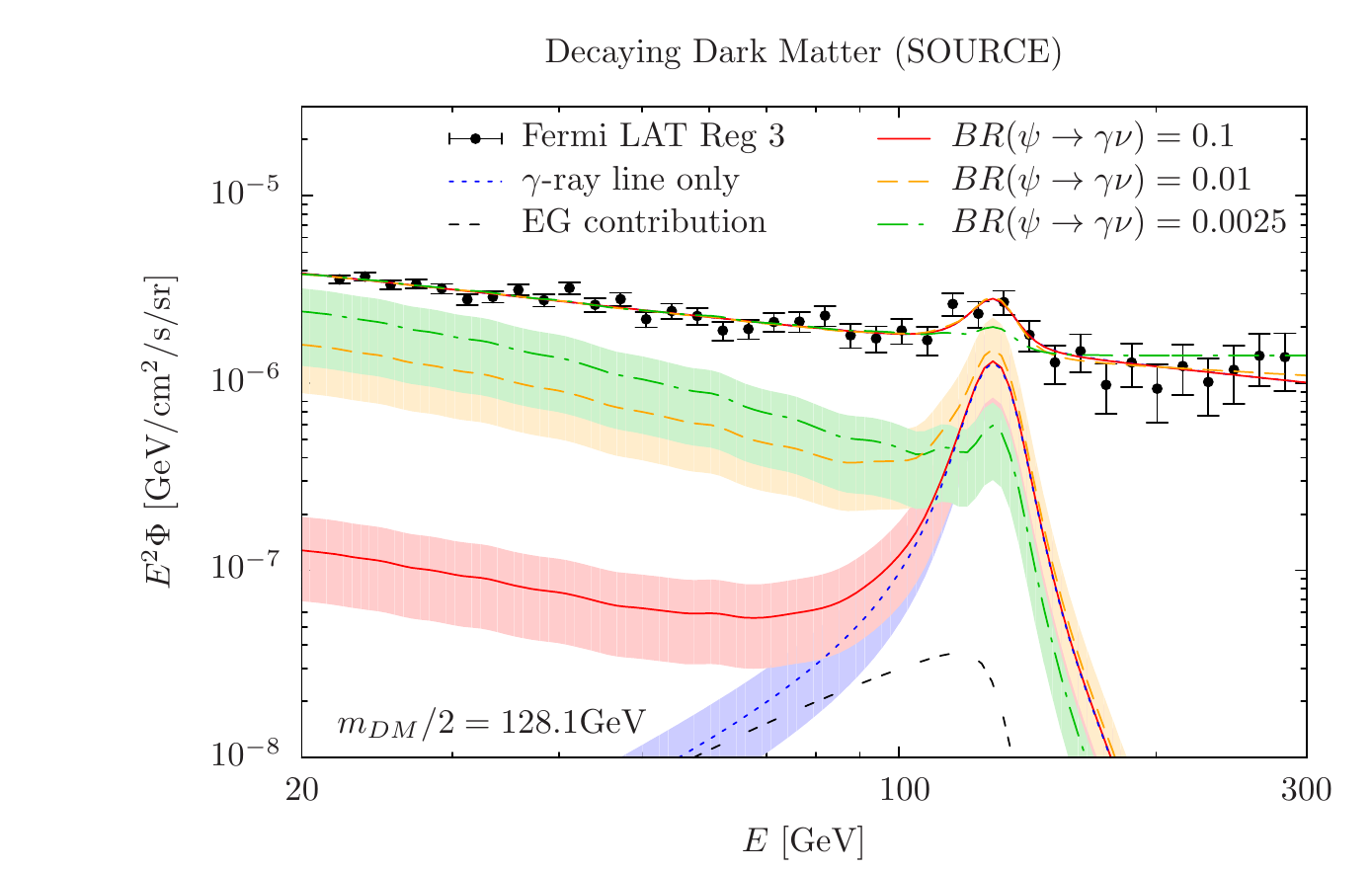}
 \\[1.5ex]
 \includegraphics[width=0.9\textwidth]{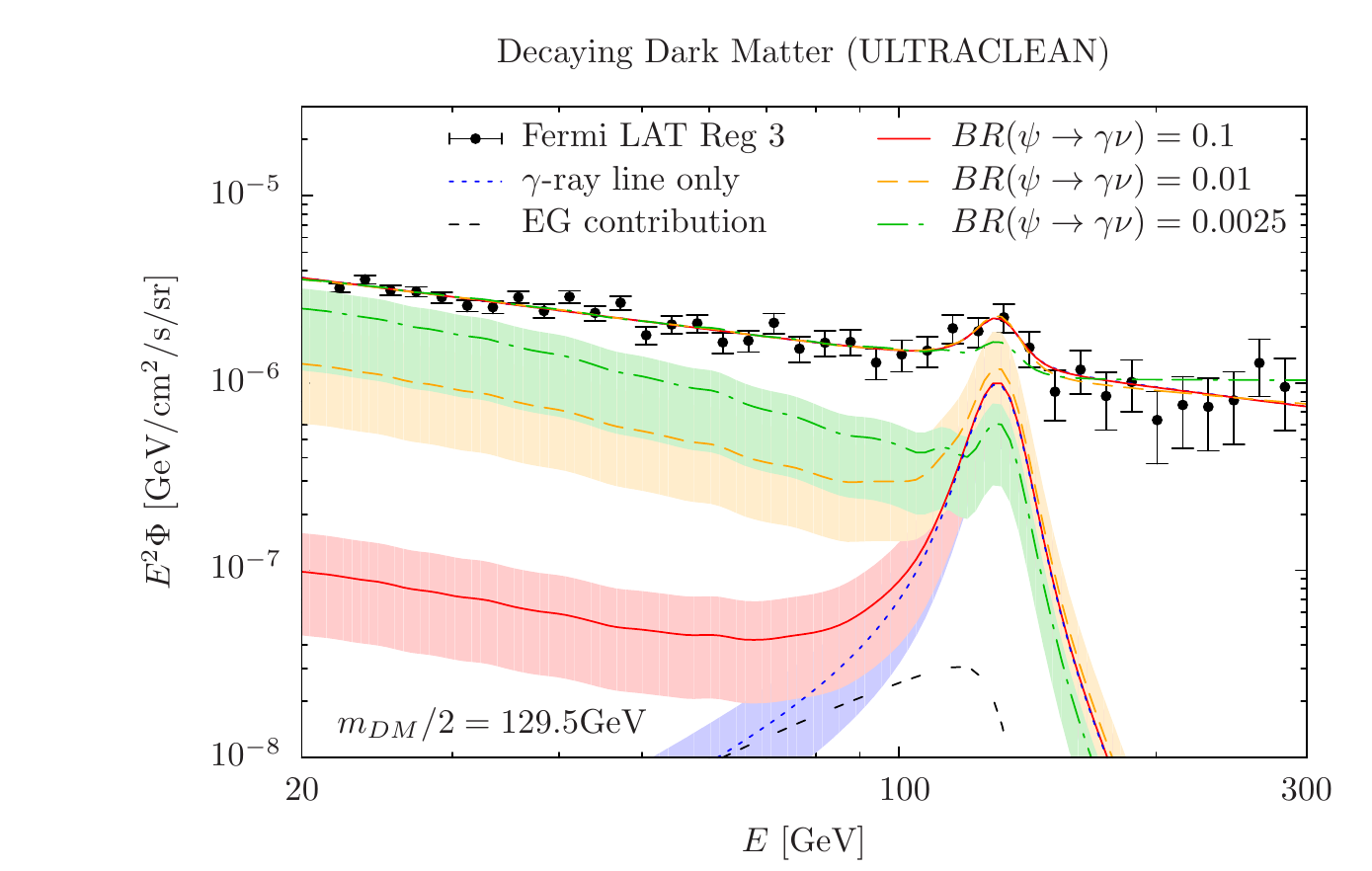}
\end{center}
 \caption{\label{fig:fluxGravitino}\small Sum of monochromatic and continuum
photon flux originating from dark matter decaying into $\gamma\nu, Z\nu, h\nu,
W^\pm\ell^\mp$ with branching ratios $\BR_\gamma=0.1$, $0.01$, and $0.0025$,
respectively. The lower lines show the contribution from dark matter decay,
and the upper lines the sum of signal and fitted power-law background. The
dashed blue line shows the flux when taking the gamma-ray line only, plus
the redshifted extragalactic contribution, into account. The gamma-ray flux
measured by Fermi-LAT corresponding to the SOURCE (upper figure) and
ULTRACLEAN (lower figure) data samples is taken from
Ref.~\cite{Weniger:2012tx}. The shaded regions correspond to $95\%$C.L.
error bands.}
\end{figure}

\begin{table}[h!]
\centering
{\scriptsize
\begin{tabular}{llc|cc|cc|cc|cc|cc}
\toprule
   \multicolumn{2}{c}{Reg}  & $E_\gamma$ &  \multicolumn{10}{c}{$\tau_{\gamma\nu}\ [10^{28}\mbox{s}]$}    \\
     &&  [\GeV] &  \multicolumn{2}{c}{$\gamma$-ray line only} & \multicolumn{2}{c}{$\BR_{\gamma\nu}=0.1$} &  \multicolumn{2}{c}{$\BR_{\gamma\nu}=0.01$} & \multicolumn{2}{c}{$\BR_{\gamma\nu}=0.0025$} & \multicolumn{2}{c}{$\BR_{\gamma\nu}=0.001$} \\
\midrule
    3 & S & 128.1 & $1.49^{+1.30}_{-0.50}$& $4.5\sigma$ & $1.46^{+1.27}_{-0.50}$& $4.5\sigma$ & $1.28^{+1.04}_{-0.41}$& $4.6\sigma$ & $3.42^{+3.26}_{-0.84}$& $3.2\sigma$ & $11.7$ & $1.0\sigma$\\[1.5ex]
    3 & UC & 129.5 & $1.94^{+2.30}_{-0.74}$& $3.8\sigma$ & $1.88^{+2.23}_{-0.71}$& $3.8\sigma$ & $1.60^{+1.73}_{-0.58}$& $4.0\sigma$ & $3.27^{+3.72}_{-0.71}$& $3.1\sigma$ & $11.4$& $0.9\sigma$\\[1.5ex]
    4 & S & 129.8 & $1.24^{+1.20}_{-0.44}$& $4.3\sigma$ & $1.21^{+1.16}_{-0.43}$& $4.3\sigma$ & $1.10^{+0.95}_{-0.35}$& $4.5\sigma$ & $3.54^{+7.86}_{-1.02}$& $2.6\sigma$ & $21.1$& $0.6\sigma$\\[1.5ex]
    4 & UC & 129.9 & $1.66^{+2.41}_{-0.67}$ & $3.5\sigma$ & $1.62^{+2.30}_{-0.65}$& $3.5\sigma$ & $1.38^{+1.70}_{-0.51}$& $3.7\sigma$ & $3.67^{+11.9}_{-1.07}$ & $2.4\sigma$ & $21.7$& $0.6\sigma$\\
\bottomrule
\end{tabular}
}
\caption{\small Partial dark matter lifetime $\tau_{\gamma\nu}$ in the
monochromatic gamma channel obtained from a fit to the Fermi-LAT data in the
energy range $20-300$~GeV in search regions 3 and 4 of
Ref.~\cite{Weniger:2012tx}, for the SOURCE (S) and ULTRACLEAN (UC) data samples,
for various fixed values of the branching ratio
$\BR_{\gamma\nu}=\tau/\tau_{\gamma\nu}$ and  using the Einasto profile. The
columns show the line energy that yields the best fit, the lifetimes with
$95\%$C.L. statistical errors, and the significance $\sigma\equiv\sqrt{TS}$
(without trial corrections) with respect to the background-only case.}
\label{tab:continuum}
\end{table}

In order to illustrate the possibilities to explain the tentative gamma-ray
line with decaying dark matter, we show in Fig.~\ref{fig:fluxGravitino}
the photon spectrum from gravitino decay, fitted to the flux measured by Fermi 
LAT within the search region 3 of Ref.~\cite{Weniger:2012tx}, for three values
of the branching ratio into monochromatic photons. The feature in the spectrum
near $130$\,GeV can be explained for branching ratios of $10\%$ and $1\%$. For
the relatively small value $0.25\%$, the continuum gamma-ray contribution
becomes so strong that it would overshoot the measured flux if one would require
the monochromatic decay mode to account for the feature in the spectrum. In this
case, the fit to the observed spectrum becomes worse, and requires a somewhat
longer total lifetime, such that the continuum flux is in accordance with the
observed one. The best fit values for the partial lifetime, as well as the
significance with respect to the background-only case obtained from search
regions 3 and 4 are shown in Tab.~\ref{tab:continuum}, for various branching
ratios. The significance of the fit remains nearly constant for branching ratios
larger than $1\%$, and then steeply decreases. For a branching ratio of $0.1\%$
or smaller, the dark matter component cannot yield a significantly better
description than the background only. 

\begin{figure}
\vspace*{-1.5cm}
\begin{center}
 \includegraphics[width=0.75\textwidth]{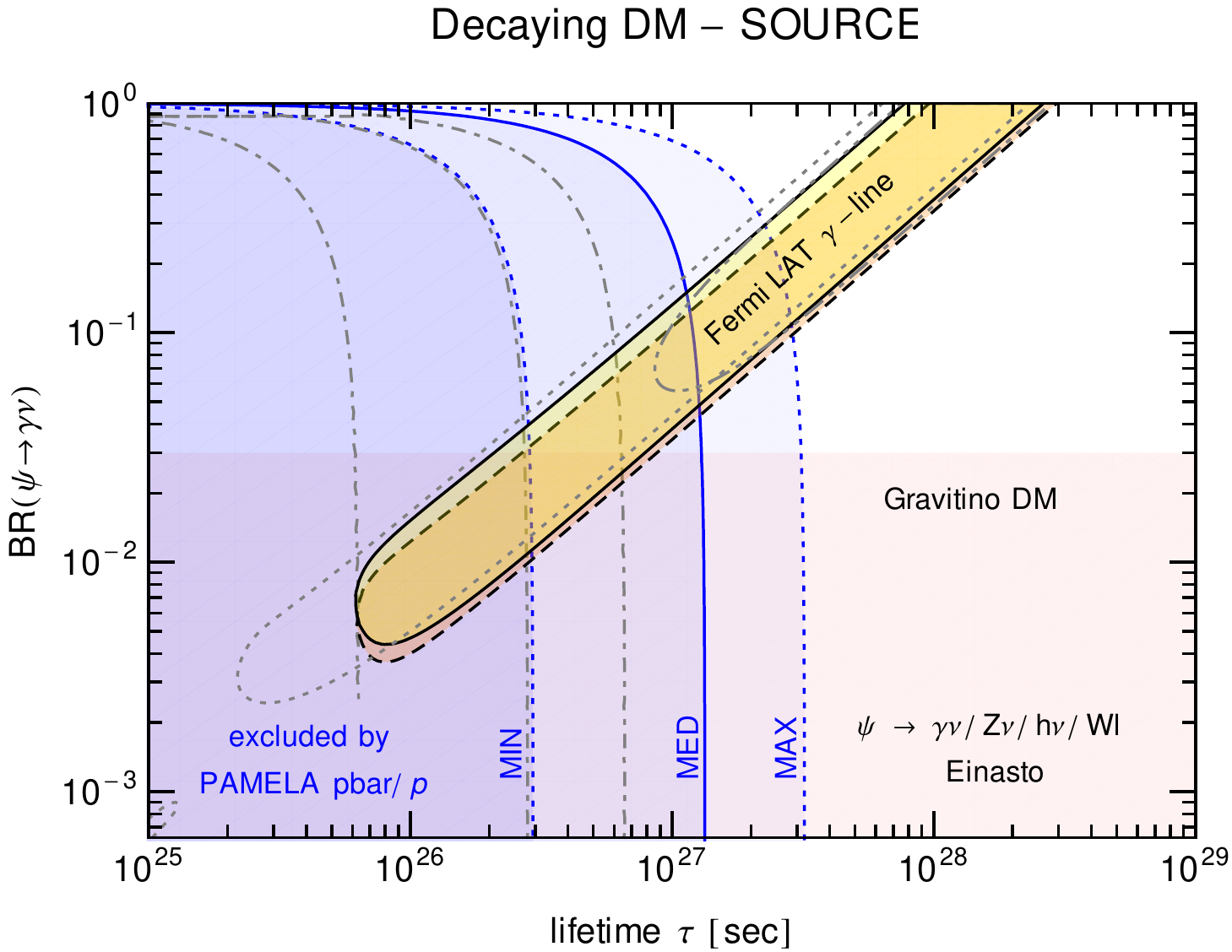}\\[1.5ex]
 \includegraphics[width=0.75\textwidth]{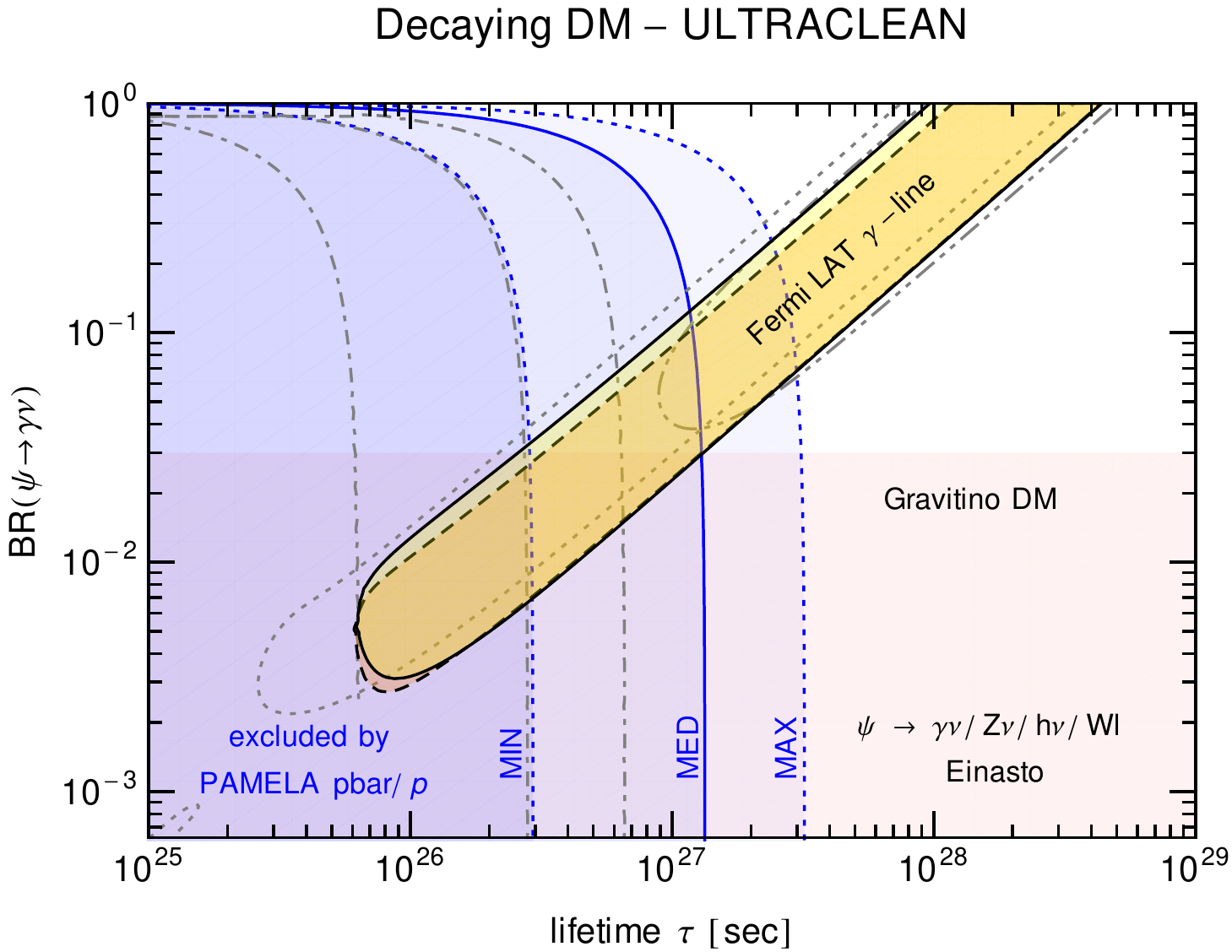}
\end{center}
 \caption{\label{fig:pbar}\small Combined fit of the monochromatic line at
$\approx 130$GeV as well as the continuum gamma-ray spectrum from dark matter
decay. The yellow and orange shaded areas are $95\%$C.L. regions for the total
dark matter lifetime $\tau$ and the branching ratio $\BR(\psi\to\gamma\nu)$
obtained from the Fermi-LAT data within search regions 3 and 4 or
Ref.~\cite{Weniger:2012tx}, respectively. The regions excluded from antiproton
constraints at $95\%$C.L. based on the PAMELA $\bar p/p$ data are shown as blue
shaded regions (see Section~\ref{sec:pbar}), as well as the expected range for
gravitino dark matter (red shaded area). Further explanations are given in the
text.}
\end{figure}

We have also performed a fit where the branching ratio is varied together with
the parameters $\alpha,\beta,\gamma$ in Eq.~(\ref{eq:templateDecayingDM}). The
resulting $95\%$C.L. contours are shown in Fig.~\ref{fig:pbar}, indicating again
that branching ratios below $0.25\%$ are disfavored because of the continuum
contribution to the photon flux. For comparison we also show the more stringent
constraints resulting from fixing the background parameters
$(\beta,\gamma)=(\beta,\gamma)|_{BR_{\gamma\nu}=1}$ to the best-fit values
obtained from the fit using a gamma-ray line only (grey dot-dashed line), as
well as the constraints obtained when using the smaller energy range
$80-200$~GeV (grey dotted line) for region 4. We also note that the results are
rather insensitive to the precise choice of the relative branching ratios among
$Z\nu,W\ell,h\nu$. 

Clearly, the limits on the continuum gamma-ray flux obtained in this way are very
conservative ones, while stronger constraints could be obtained, {\it e.g.}, by
taking photons produced via inverse Compton scattering into account
({\it e.g.}~\cite{Cirelli:2012ut}) or extending the energy range below $20$~GeV.
However, this would also make the conclusions more dependent on details of the
background spectrum and therefore on the astrophysical processes close to the
Galactic center. In this sense, the limits presented here may be regarded as a
rather robust consistency check, based on the assumption that the background can
be modeled by a power law in the energy range $20-300$~GeV.

\subsection{Annihilating dark matter}

One of the most studied candidates for dark matter is the lightest neutralino
within the MSSM, for which the annihilation cross sections into $\gamma\gamma$
and $\gamma Z$, that yield monochromatic gamma rays, are loop
suppressed~\cite{Bergstrom:1997fh}. Consequently, the branching ratios into
these final states are typically at the permille level, and the continuum flux
dominates the gamma spectrum\footnote{In some extensions of the Standard Model DM
annihilation can also produce intense gamma-ray lines. For recent examples and
references see~\cite{Weniger:2012tx,Dudas:2012pb,Jackson:2009kg}.}. 

In the following we perform first a model independent analysis in order to
determine the requirements on the branching ratio into monochromatic photons,
and then check whether some well-motivated scenarios are compatible with these
constraints.

\begin{figure}
\vspace*{-1.5cm}
\begin{center}
 \includegraphics[width=0.9\textwidth]{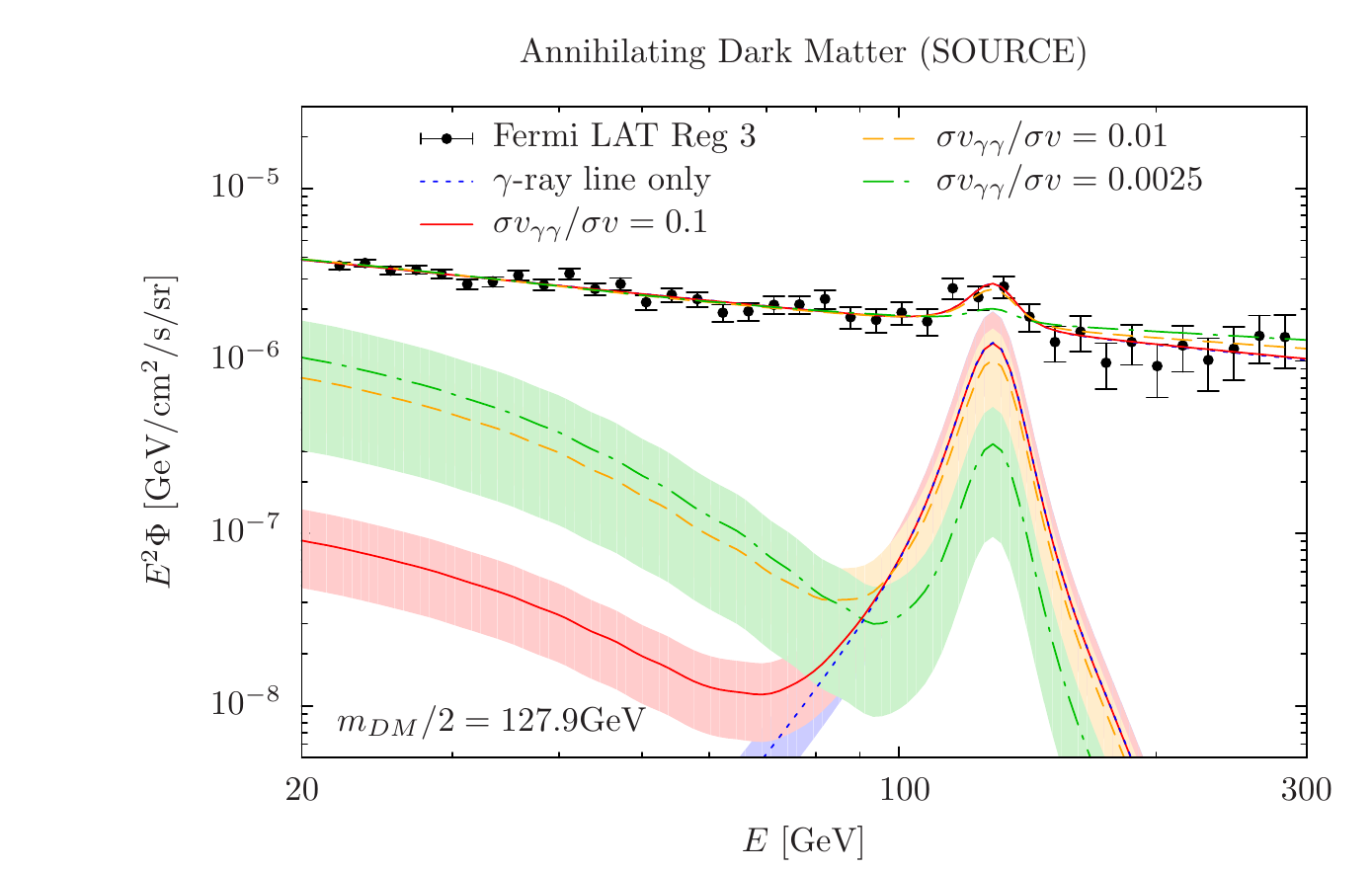}
 \\[1.5ex]
 \includegraphics[width=0.9\textwidth]{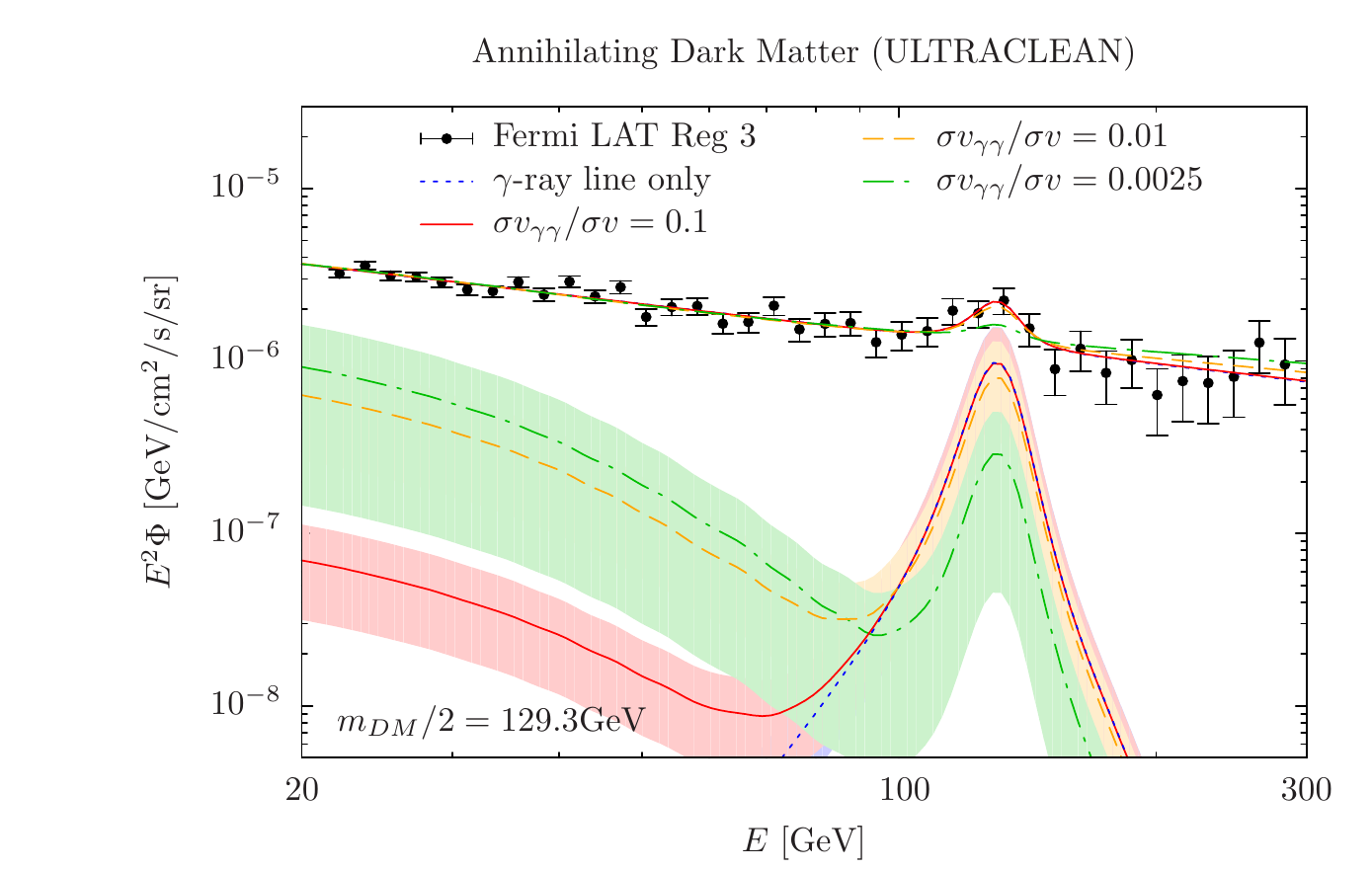}
\end{center}
 \caption{\label{fig:fluxWWgamgam}\small Sum of monochromatic and continuum
photon flux originating from dark matter annihilating into $\gamma\gamma$ and
$W^+W^-$ with branching ratios $\BR_{\gamma}=0.1$, $0.01$, and $0.0025$,
respectively. The lower lines show the contribution from dark matter
annihilation, and the upper lines the sum of signal and fitted power-law
background. The dashed blue line shows the flux when taking the line only into
account. The Fermi data are the same as in Fig.~\ref{fig:fluxGravitino}.}
\end{figure}

\begin{figure}
\vspace*{-1.5cm}
\begin{center}
 \includegraphics[width=0.75\textwidth]{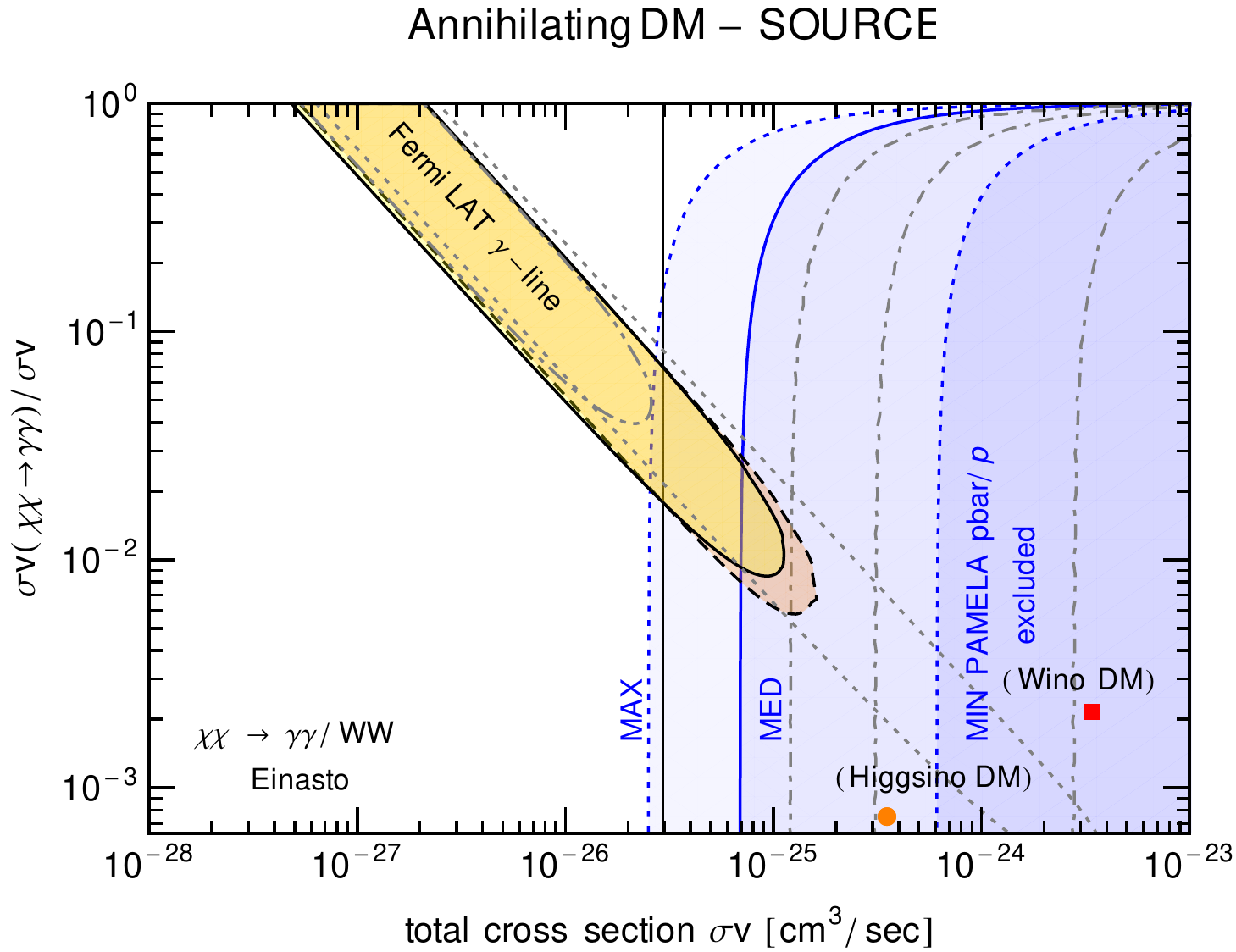}\\[1.5ex]
 \includegraphics[width=0.75\textwidth]{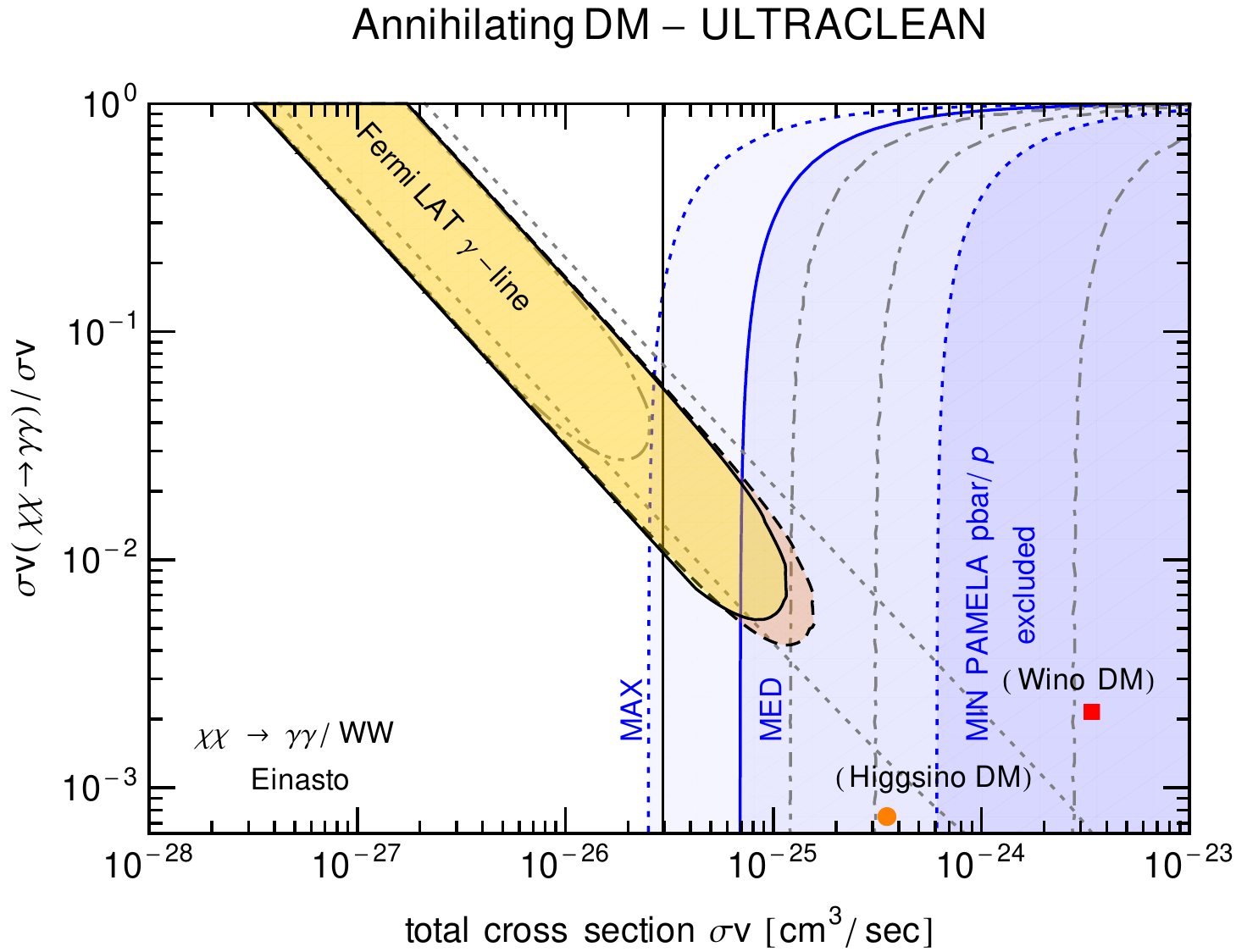}
\end{center}
 \caption{\label{fig:pbarWW}\small Combined fit of the monochromatic line at
$\approx 130$GeV as well as the continuum gamma-ray spectrum from dark matter
annihilation into $\gamma\gamma$ and $W^+W^-$ final states. The shaded areas
correspond to $95\%$C.L. regions (yellow/orange: Fermi-LAT Reg 3/4 best-fit
region; blue: PAMELA $\bar p/p$ excluded region, see Section~\ref{sec:pbar})
for the total annihilation cross section $\sigma v$ and the branching ratio
$\sigma v_{\gamma\gamma}/\sigma v$. The prediction from higgsino and wino dark
matter (red and orange dots) is shown for illustration, as well as the thermal
cross-section (straight black line).}
\end{figure}

In particular, we first assume that dark matter annihilates into $\gamma\gamma$
and $W^+W^-$ final states, giving rise to a line at $E_\gamma=m_{DM}$ and a
continuum photon flux in the energy range $E<m_{DM}$. The resulting photon flux
is shown in Fig.~\ref{fig:fluxWWgamgam}, for several values of the branching
ratio $\BR_{\gamma}=\sigma v_{\gamma\gamma}/(\sigma v_{\gamma\gamma}+\sigma v_{WW})$.
Similarly to decaying dark matter, we find that the feature in the Fermi
spectrum at $\approx 130$\,GeV can be well fitted for branching ratios $10\%$
or $1\%$, while the continuum spectrum becomes dominant for $0.25\%$. Note that
the photon spectrum exhibits a more pronounced dip between the continuum and
monochromatic contributions as compared to decaying dark matter, due to the
absence of the extragalactic contribution. The statistical significance and
best-fit values of the cross section are given in Tab.~\ref{tab:continuumWWgamgam}
for various branching ratios, and exhibit a similar dependence than for
decaying dark matter. One might notice that the significance of the fit always
decreases when lowering $\BR_\gamma$, while for decaying dark matter it is
slightly better for $\BR_\gamma\sim 1\%$. However, these differences are too
small to be taken seriously. A common feature is the drastic decrease in the
significance for branching ratios smaller than $1\%$. As before, we also
performed a fit with variable branching ratio. The resulting $95\%$C.L. regions
are shown in Fig.~\ref{fig:pbarWW}. Depending on the search region and on the
data sample used for the analysis, we find that a branching ratio larger than
$0.4-0.8\%$ is required. For comparison, the constraints obtained when using a
fixed background contribution (grey dot-dashed line) or when using the smaller
energy range $80-200$\,GeV (grey dotted lines) are shown as well. The sensitivity
to the continuum spectrum is essentially lost in the latter case for annihilating
dark matter, due to the drop of the flux below the line. On the other hand, we
checked that removing the lower limit $\gamma>2.0$ on the background slope does
not affect the confidence regions.

\begin{table}[t]
\centering
{\scriptsize
\begin{tabular}{llc|cc|cc|cc|cc|cc}
\toprule
   \multicolumn{2}{c}{Reg}  & $E_\gamma$ &  \multicolumn{10}{c}{$\sigma v_{\gamma\gamma}\ [10^{-27}\mbox{cm$^3/$s}]$}    \\
     &&  [\GeV] &  \multicolumn{2}{c}{$\gamma$-ray line only} & \multicolumn{2}{c}{$\BR_{\gamma\gamma}=0.1$} &  \multicolumn{2}{c}{$\BR_{\gamma\gamma}=0.01$} & \multicolumn{2}{c}{$\BR_{\gamma\gamma}=0.0025$} & \multicolumn{2}{c}{$\BR_{\gamma\gamma}=0.001$} \\
\midrule
    3 & S & 127.9 & $1.25^{+0.65}_{-0.58}$& $4.5\sigma$ & $1.24^{+0.64}_{-0.58}$& $4.5\sigma$ & $0.99^{+0.52}_{-0.49}$& $4.1\sigma$ & $0.32^{+0.21}_{-0.23}$& $2.7\sigma$ & $0.10$ & $1.7\sigma$\\[1.5ex]
    3 & UC & 129.3 & $0.98^{+0.64}_{-0.54}$& $3.8\sigma$ & $0.97^{+0.60}_{-0.53}$& $3.8\sigma$ & $0.80^{+0.51}_{-0.47}$& $3.5\sigma$ & $0.29^{+0.22}_{-0.25}$& $2.3\sigma$ & $0.09$& $1.3\sigma$\\[1.5ex]
    4 & S & 129.6 & $1.24^{+0.68}_{-0.62}$& $4.3\sigma$ & $1.21^{+0.68}_{-0.61}$& $4.2\sigma$ & $0.86^{+0.51}_{-0.48}$& $3.6\sigma$ & $0.22$& $1.9\sigma$ & $0.05$ & $0.8\sigma$\\[1.5ex]
    4 & UC & 129.8 & $0.93^{+0.63}_{-0.55}$& $3.5\sigma$ & $0.92^{+0.63}_{-0.55}$& $3.5\sigma$ & $0.68^{+0.50}_{-0.46}$& $3.0\sigma$ & $0.19$& $1.6\sigma$ & $0.04$& $0.6\sigma$\\[1.5ex]
\bottomrule
\end{tabular}
}
\caption{\small Annihilation cross section in the monochromatic gamma channel
obtained from a fit to the Fermi-LAT data in the energy range $20-300$GeV in
search regions 3 and 4 of Ref.~\cite{Weniger:2012tx}, for the SOURCE (S) and
ULTRACLEAN (UC) data samples, for various fixed values of the branching ratio
$\BR_{\gamma\gamma}=\sigma v_{\gamma\gamma}/(\sigma v_{\gamma\gamma}+\sigma v_{WW})$,
and  using the Einasto profile. The columns show the line energy that yields
the best fit, the cross sections with $95\%$C.L. statistical errors, and the
significance $\sigma\equiv\sqrt{TS}$ (without trial corrections) with respect
to the background-only case.  The small differences in the statistical
significances between dark matter decay (see Tab.~\ref{tab:continuum}) and
annihilation are mainly due to the extragalactic contribution as well as a
slightly different continuum spectrum.} 
\label{tab:continuumWWgamgam}
\end{table}

\begin{figure}
\vspace*{-1.5cm}
\begin{center}
 \includegraphics[width=0.75\textwidth]{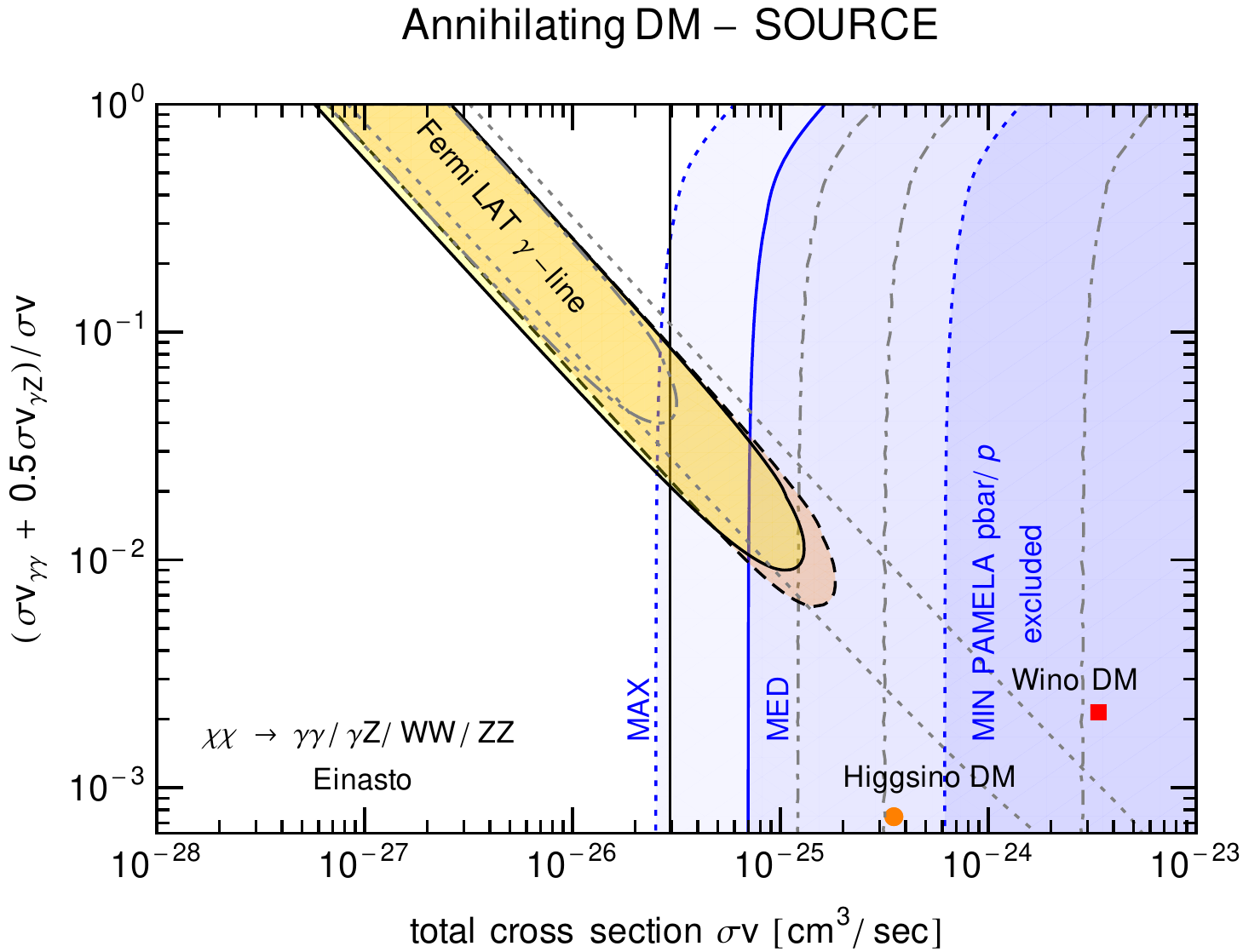}\\[1.5ex]
 \includegraphics[width=0.75\textwidth]{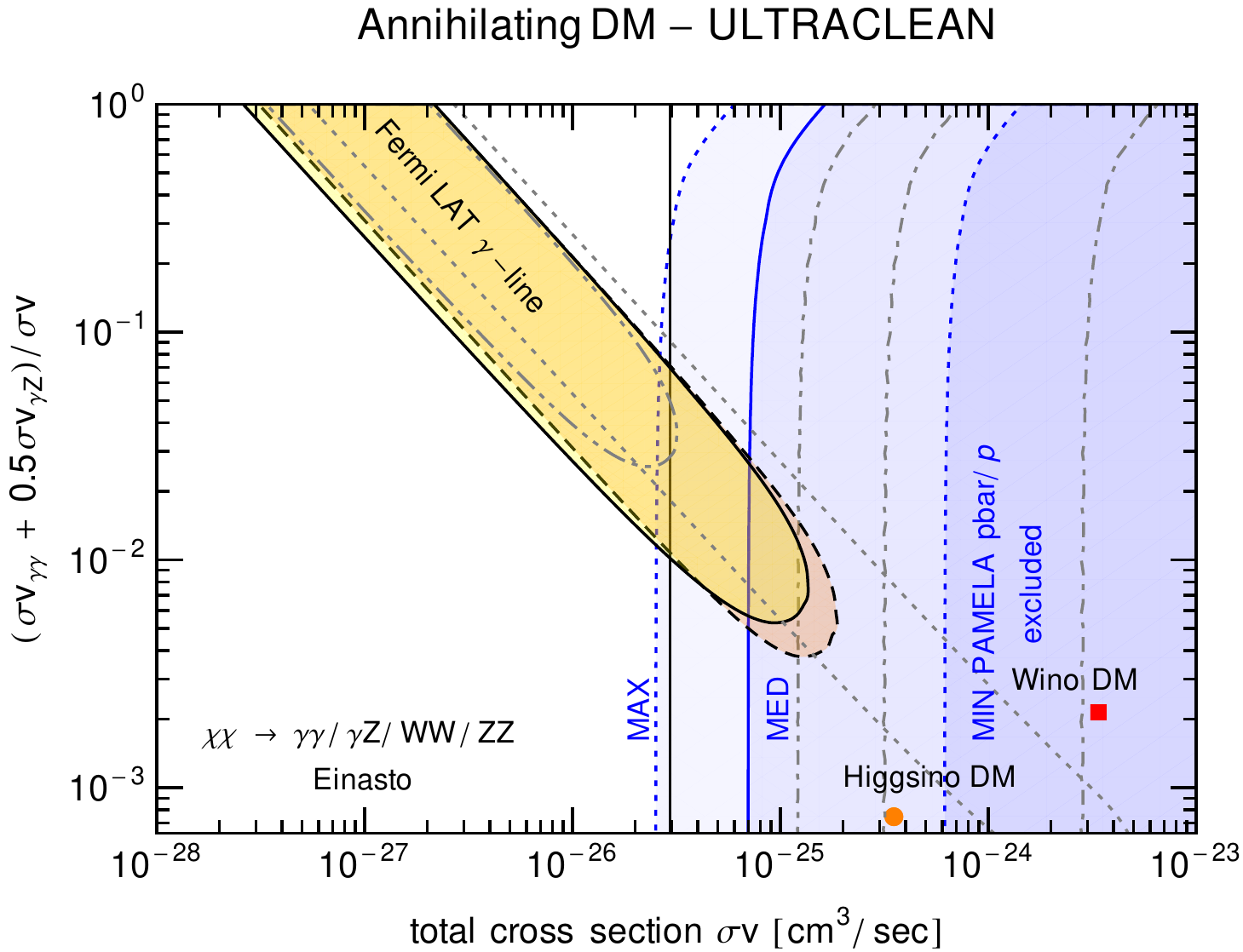}
\end{center}
 \caption{\label{fig:pbarHiggsino}\small As Fig.\,\ref{fig:pbarWW}, but for a
model with two lines from annihilation into $\gamma\gamma$ and $\gamma Z$, as
well as continuum photons from annihilations into $W^+W^-$ and $ZZ$ final
states. The axes correspond to the total cross section $\sigma v$ and to the
weighted ratio $(\sigma v_{\gamma\gamma}+0.5\sigma v_{\gamma Z})/\sigma v$,
respectively. For the fit we fixed the ratios
$\sigma v_{\gamma Z}/(2\sigma v_{\gamma\gamma})=1.68$ and
$\sigma v_{ZZ}/\sigma v_{WW}=0.66$ motivated by the higgsino dark matter
scenario. The actual prediction from higgsino and wino dark matter is also
shown (red and orange dots). PAMELA antiproton constraints are discussed in
Section~\ref{sec:pbar}.}
\end{figure}

As a second generic example, we consider dark matter annihilating into $\gamma Z$
and $\gamma\gamma$, giving rise to two lines at $E_{\gamma,1}=m_{DM}$ and
$E_{\gamma,2}=m_{DM}-M_Z^2/(4 m_{DM})$. In addition, annihilations into $W^+W^-$
and $ZZ$ yield a continuum spectrum. Motivated by higgsino dark matter, we fix
the relative contribution of the two lines according to
$\sigma v_{\gamma Z}/(2\sigma v_{\gamma\gamma})=1.68$ as well as
$\sigma v_{ZZ}/\sigma v_{WW}=0.66$. However, we freely vary the weighted
branching ratio
$\BR_\gamma=(\sigma v_{\gamma\gamma}+0.5\sigma v_{\gamma Z})/\sigma v_{tot}$. In
this case we obtain the best fit for dark matter masses in the range
$132-137$\,GeV, depending on the search region and the data sample, and with
similar significance than for one line. The resulting $95\%$C.L. regions are
shown in Fig.~\ref{fig:pbarHiggsino}, which are very similar to the case of a
single line discussed before, indicating that the lower limit
$\BR_\gamma \gtrsim 0.4-0.8\%$ is rather insensitive to the detailed assumptions
about the cross sections of the various annihilation channels.

\begin{figure}
\vspace*{-1.5cm}
\begin{center}
 \includegraphics[width=0.9\textwidth]{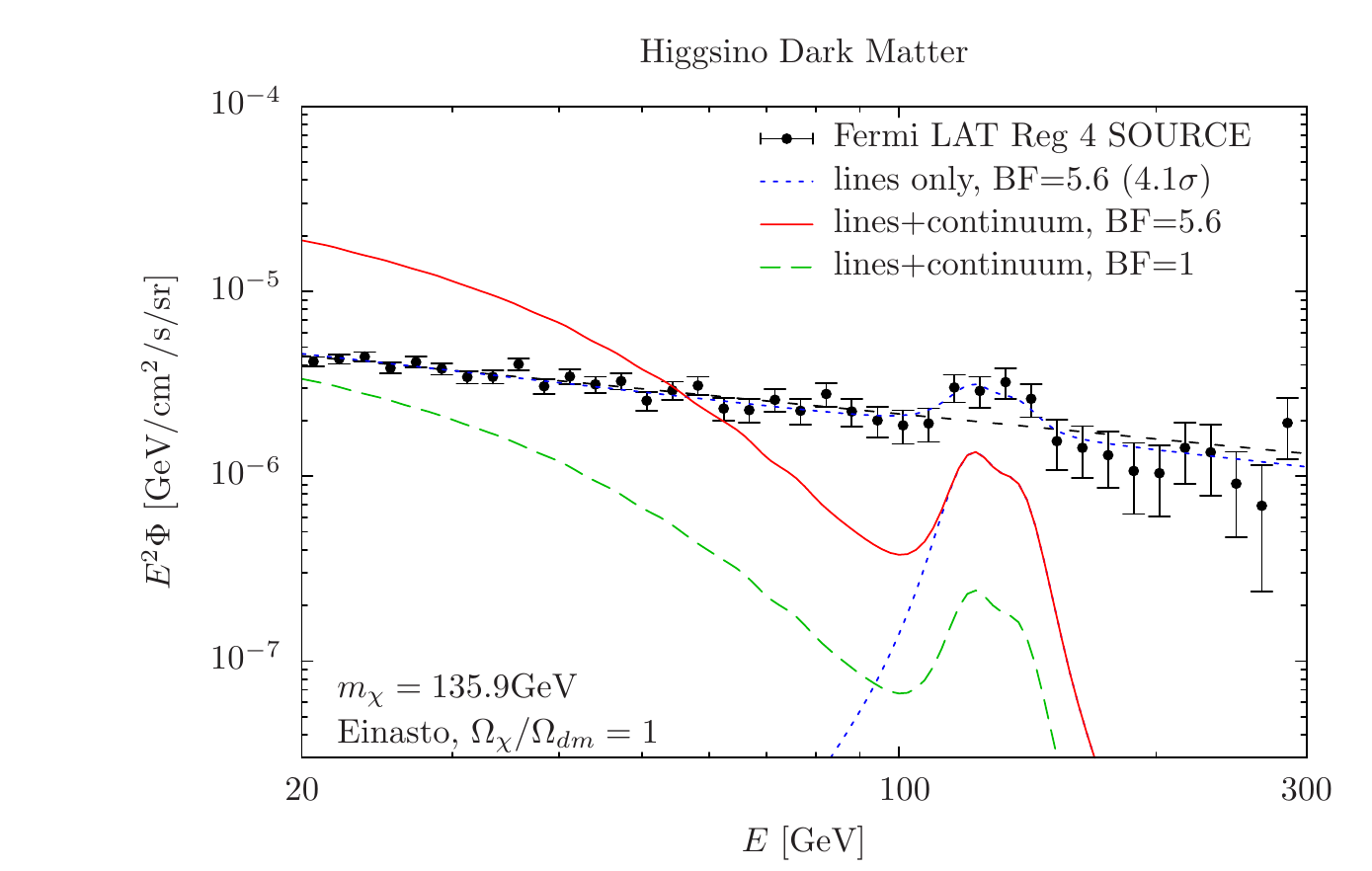}
 \\[1.5ex]
 \includegraphics[width=0.9\textwidth]{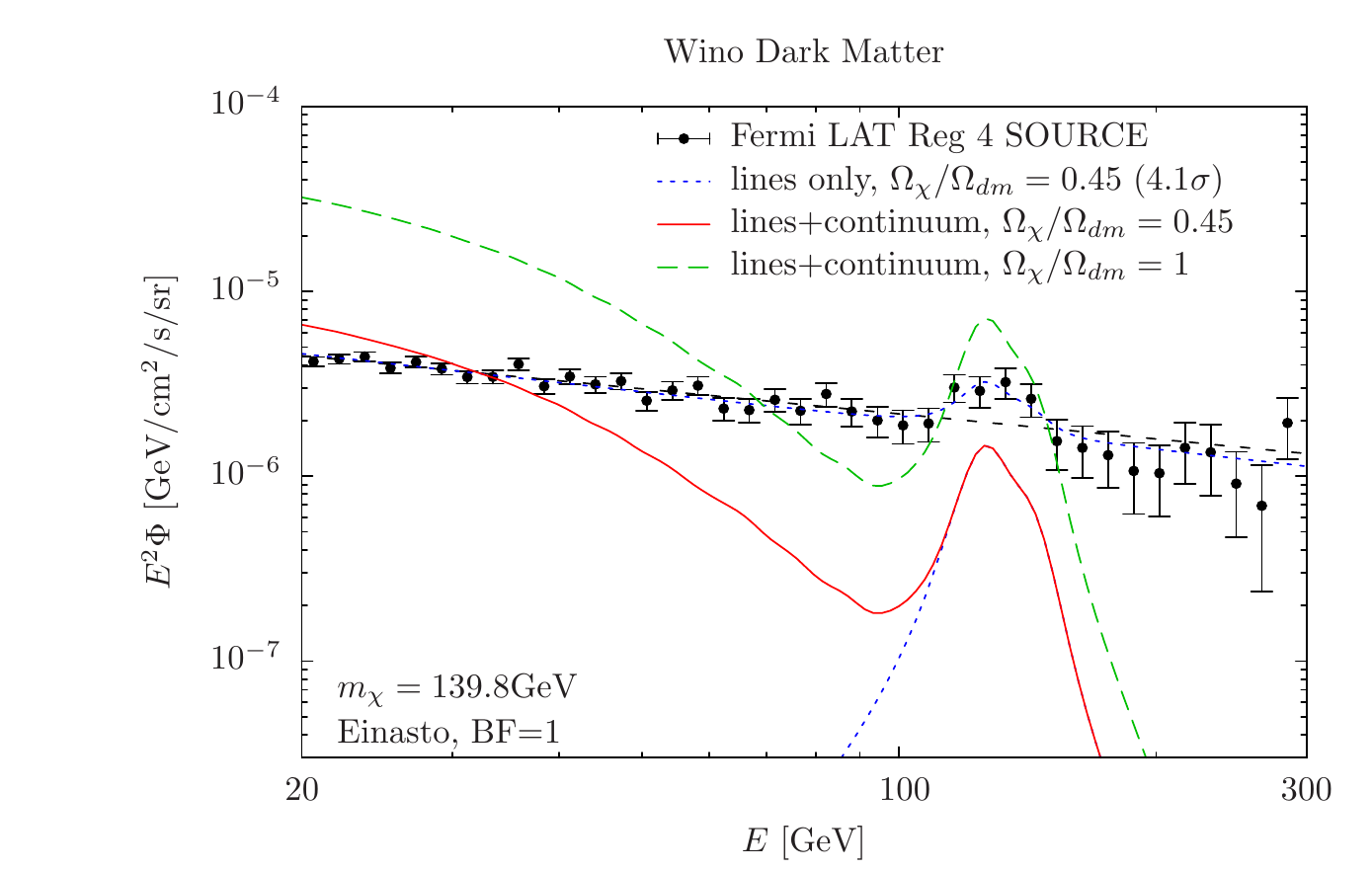}
\end{center}
 \caption{\label{fig:fluxHiggsino}\small Predicted gamma-ray flux for higgsino
dark matter (upper graph) and wino dark matter (lower graph, green dashed lines).
The monochromatic gamma-ray flux is too small(large) in the higgsino(wino) case
for explaining the Fermi excess at $\approx 130$\,GeV when assuming an Einasto
profile. We also show the flux obtained for a boost factor $BF=5.6$ in the
higgsino case and when assuming that the wino accounts only for $45\%$ of the
dark matter density, respectively (red solid lines). In both cases, the
monochromatic contribution to the flux would yield a good fit to the Fermi data
(blue dotted lines). However, the continuum contribution to the gamma-ray flux
overshoots the measured flux at lower energies, and therefore both scenarios can
be ruled out. The Fermi data are taken from~\cite{Weniger:2012tx}.}
\end{figure}

Finally, we briefly remark that both higgsino and wino-like dark matter can be
ruled out as an explanation for the tentative gamma line based on the large
continuum photon flux obtained in both cases. The cross sections for two
representative scenarios are given in Tab.~\ref{tab:scenario}, leading to
$\BR_\gamma\sim 0.08\%(0.2\%)$ for higgsino(wino)-like dark matter. Even when
allowing for a boost factor or a mixed dark matter scenario in order to achieve
a line flux of the required size, the continuum flux overshoots the measured
flux at energies around $20$\,GeV, which is illustrated in
Fig.~\ref{fig:fluxHiggsino}. In particular, we remark that this conclusion is
independent of the dark matter distribution, and therefore of the uncertainties
associated with it. According to the recent analysis in \cite{Hryczuk:2011vi},
electroweak corrections could even further reduce the $\gamma \gamma$ and
$\gamma Z$ cross sections for low-mass wino dark matter by a sizeable amount.

\begin{table}[t]
\centering
\begin{tabular}{ccc|ccc|cc}
\toprule
  & $\mu$ & $M_2$ &  $m_{\chi_0^1}$ & $m_{\chi_0^2}$ & $m_{\chi_\pm^1}$ & $\sigma v_{\gamma\gamma}(\sigma v_{\gamma Z})$ & $\sigma v_{WW}(\sigma v_{ZZ})$ \\
\midrule
 H & 139 & $1000$ &  135.89 & 144.44 & 139.20 & $1.0(3.4)\cdot 10^{-28}$ & $2.1(1.4)\cdot 10^{-25}$  \\ 
 W & 400 & 143 &  139.79 & 408.08 & 139.94 & $2.0(10.9)\cdot 10^{-27}$ & $3.4(0.0)\cdot 10^{-24}$  \\ 
\bottomrule
\end{tabular}
\caption{\small Mass spectrum, annihilation cross sections and defining parameters
for two scenarios with higgsino (H) and wino-like (W) LSP. Particle masses are
given in GeV, and cross sections in $\text{cm}^3/\text{s}$. The cross sections
are rather insensitive to the other parameters, which we fixed to $A=0$,
$M_{1,3}=\tilde m = 1$\,TeV and $\tan\beta=7$ for illustration. For the wino
case the $\mu$ parameter was chosen such that the neutralino-chargino mass
splitting is of the same order than the minimal splitting expected from loop
corrections~\cite{Cirelli:2005uq} in order to yield an estimate for the maximal
possible gamma cross sections. 
}
\label{tab:scenario}
\end{table}

\section{Constraints from antiprotons}\label{sec:pbar}

Constraints that are complementary to the continuum gamma-ray spectrum arise from
the production of antiprotons in the decay and fragmentation of the weak gauge
bosons, Higgs bosons or quarks produced in dark matter annihilation or decay. The
resulting primary flux of antiprotons can be potentially observed in the cosmic
ray spectrum measured at the Earth. Antiprotons are produced with a rate per
unit of kinetic energy and unit volume at a position $r$ with respect to the
center of the Galaxy given by
\begin{equation}
  Q_{\bar{p}}(E,r)=\left\{ \frac{\rho_{dm}(r)}{m_{DM}}\frac{1}{\tau} \atop \frac{\rho_{dm}(r)^2}{2m_{DM}^2}\,\sigma v_{tot} \right\}\times \sum_f \BR_f \frac{dN_f^{\bar{p}}}{dE}\quad \mbox{for} \quad \left\{\rule{0mm}{5mm}\ { \mbox{decay}\rule{12mm}{0mm}} \atop { \mbox{annihilation}\rule{0mm}{5mm} }  \right. \;,
  \label{eqn:source-term}
\end{equation}
where $dN_f^{\bar p}/dE$ is the number of antiprotons per kinetic energy and per
decay produced in the annihilation/decay channels $f$ with branching ratio given
by $\BR_f$, and $\rho_{dm}(r)$ is the dark matter density profile.

The antiproton-to-proton flux ratio observed by the PAMELA satellite experiment
\cite{Adriani:2010rc} agrees well with the expectation for the secondary
antiproton spectrum produced by spallation of ordinary cosmic rays on the
interstellar medium. Therefore, upper limits on a possible primary component
from dark matter annihilation or decay can be derived (see
{\it e.g.}~\cite{Donato:2008jk, Ibarra:2008qg, Buchmuller:2009xv,Garny:2011cj,
Evoli:2011id,Garny:2012}). The main source of uncertainty enters due to the
propagation of antiprotons in the Galactic magnetic field from the production
point to the Earth. In order to obtain lower limits on the dark matter lifetime
and upper limits on the annihilation cross section we follow here the approach
described in detail in Refs.~\cite{Garny:2012,Garny:2011cj}. In particular, we
describe the propagation by a cylindrical two-zone diffusion model using three
representative sets of parameters leading to minimum, medium and maximum
antiproton fluxes consistent with the charged cosmic ray
abundances~\cite{Bringmann:2006im}. In order to obtain conservative limits we
use the minimal secondary antiproton spectrum from
Refs.~\cite{Donato:2001ms,Garny:2011cj}, and use a lower cut of $T>1.5$\,GeV for
the kinetic energy.

For the case of dark matter decaying into $Z\nu,h\nu,W\ell$ with the same
branching ratios used above, the $95\%$C.L. lower limits on the dark matter
lifetime is also shown in Fig.~\ref{fig:pbar} as blue shaded regions, for the
three sets of propagation parameters and the Einasto dark matter profile. For
a branching ratio much smaller than one, the $95\%$C.L. lower limit on the
lifetime is given by $\tau \gtrsim (0.29,1.34,3.31)\times 10^{27}$s for the
minimum, medium and maximum propagation models, respectively. Compared to the
uncertainty arising from the propagation model, the dependence of the lower
limit on the dark matter profile is rather weak. As can be seen in
Fig.~\ref{fig:pbar}, the constraints from antiprotons require the branching
ratio to lie above $\sim\mathcal{O}(1\%,5\%,10\%)$, depending on the propagation
model, if we require that the partial lifetime into monochromatic photons
accounts for the gamma feature at $130$\GeV. Even for the minimal propagation
model, these lower limits on the branching ratio are more stringent than the
ones obtained from the continuum gamma-ray flux. However, they are also more
dependent on astrophysical uncertainties related to propagation as well as the
secondary antiproton flux~\cite{Evoli:2011id}. For comparison, we also show the
lower limits on the lifetime obtained from just requiring that the primary
antiproton flux does not overshoot the observed antiproton-to-proton ratio. The
resulting limits are somewhat weaker, $\tau\gtrsim (0.64,2.81,6.62)\times 10^{26}$s.
For the most conservative case, {\it i.e.} for minimal propagation, the resulting
lower limit on the branching ratio is comparable to the one obtained from the
continuum gamma-ray flux (see Fig.~\ref{fig:pbar}).

For the case of annihilating dark matter into $W^+W^-$, the antiproton
constraints shown in Fig.~\ref{fig:pbarWW} correspond to upper limits on the
annihilation cross section $\sigma v \lesssim (6.1,0.69,0.25)\times
10^{-25}$cm$^3/$s when taking the secondary flux into account. The resulting
lower limits on the branching ratio $\sigma v_{\gamma\gamma}/\sigma v$ are
more stringent or comparable to the ones from the continuum flux for maximum
or medium propagation, respectively, as can be seen in Fig~\ref{fig:pbarWW}.
In the conservative case, when taking only the primary flux into account,
one obtains rather weak limits $\sigma v \lesssim (2.8,0.31,0.12)\times
10^{-24}$cm$^3/$s that are comparable to the continuum flux only in the
maximum case. The antiproton constraints are very similar for the case when
dark matter annihilates into $\gamma\gamma/\gamma Z/WW/ZZ$ final states
(see Fig.~\ref{fig:pbarHiggsino}), except for the antiprotons from $\gamma Z$
that contribute for very large branching ratios.

Altogether, we find that antiproton constraints are typically the leading ones
for decaying dark matter, while the continuum gamma-ray flux is more or equally
important for annihilating dark matter.

\section{Morphology of the Fermi excess at 130 GeV}\label{sec:Profile}

Apart from constraints arising from the gamma-ray spectrum or from antiprotons,
the Fermi-LAT data also provide valuable information about the spatial
distribution of the gamma-ray flux observed towards the Galactic center (the
angular resolution of the LAT is $\Delta\theta\sim 0.2^\circ$~\cite{Weniger:2012tx}).
The hints for a feature in the spectrum are based on about 69(57) photons
collected within 43 months inside search regions 3(4) (solid angle
$\Delta\Omega/4\pi=0.025(0.015)$) defined in~\cite{Weniger:2012tx} and the SOURCE
sample, and 53(46) for the ULTRACLEAN sample. Consequently, any information
about the detailed spatial distribution of the excess within these regions is
at present necessarily limited by statistics. Nevertheless, the amount of data
allows to test the compatibility with certain spatial distributions. In
particular, in the analysis of \cite{Su:2012ft} it was found that using a
Gaussian template the excess appears to originate from a rather small region
$\approx 3^\circ$\,FWHM, possibly with a small offset from the Galactic center of
$1.5^\circ$ along the Galactic plane. Additionally, it was argued that an Einasto
or contracted NFW profile yields an even better description, when also taking an
offset into account.

\begin{figure}
\hspace*{-0.25cm}
\begin{tabular}{ll}
 \includegraphics[width=0.99\textwidth]{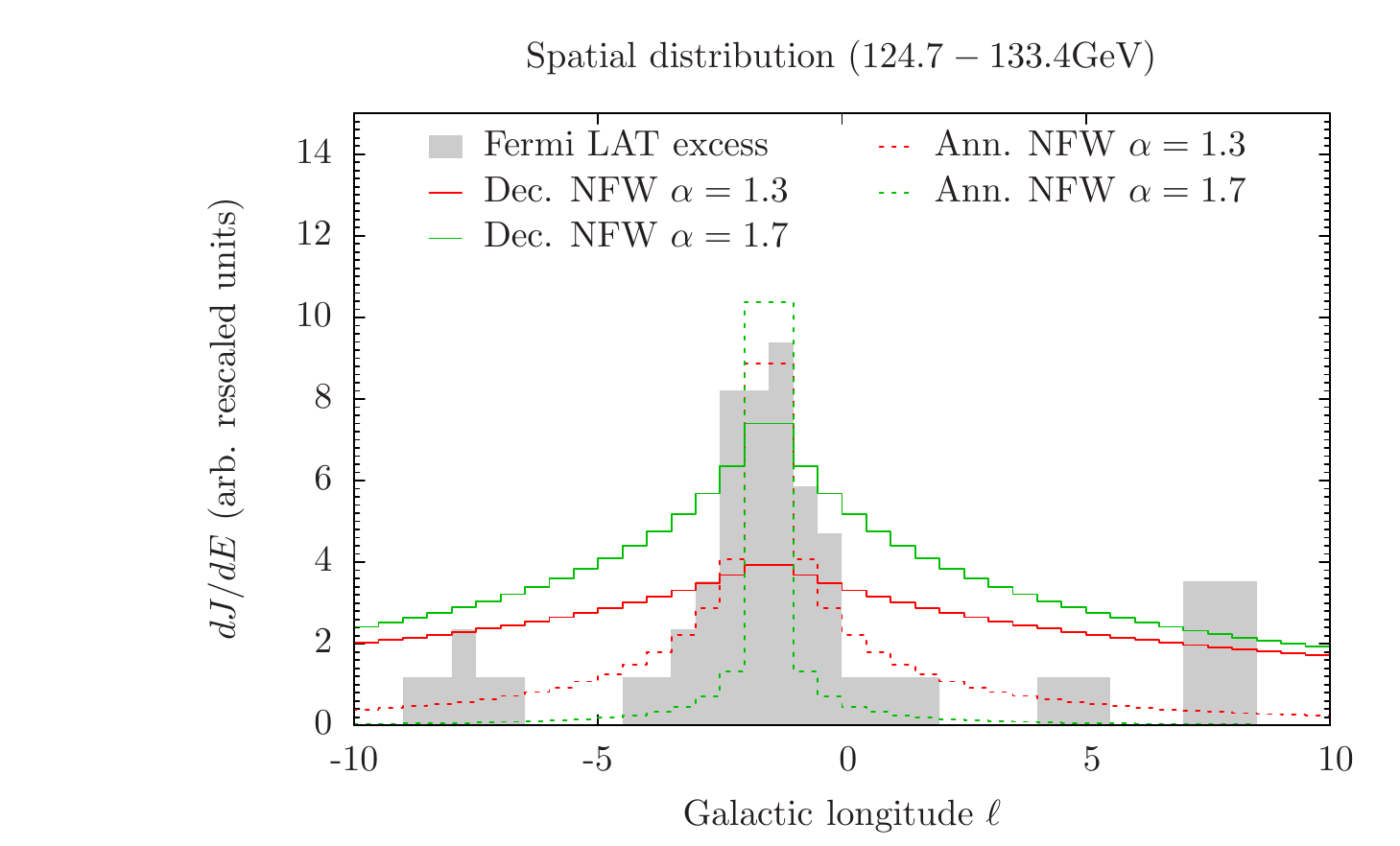}\\
 \includegraphics[width=0.99\textwidth]{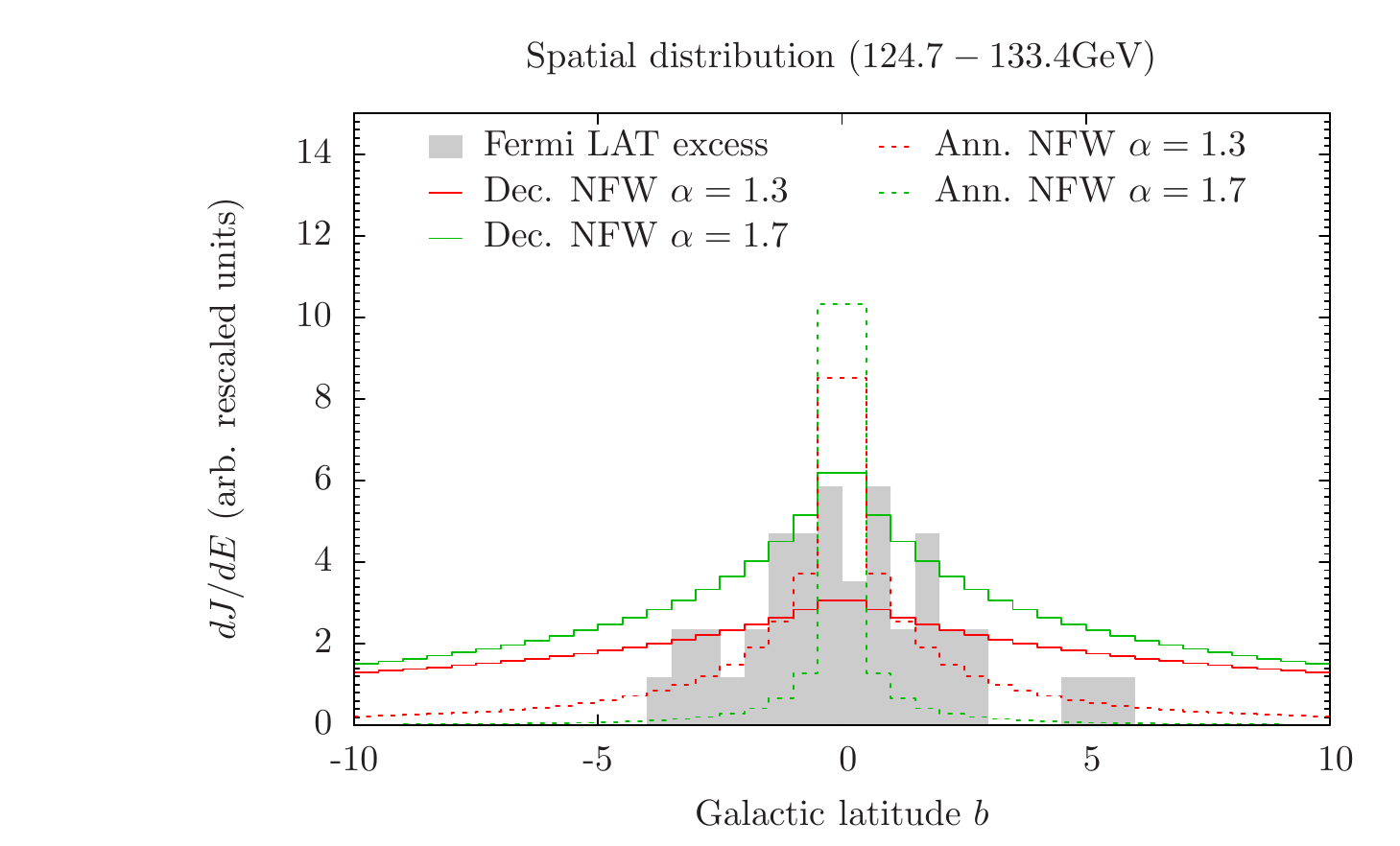}
\end{tabular}
 \caption{\small Expected spatial profile of the gamma-ray emission from sections
along the Galactic plane ($\Delta\ell=0.5^\circ, |b|<5^\circ$, upper graph) and
perpendicular to the plane ($\Delta b=0.5^\circ, -5^\circ<\ell<-2^\circ$, lower
graph) for decaying and annihilating dark matter and two density profiles centered
around $(\ell,b)=(-1.5^\circ,0)$. For illustration, we also show the count maps
indicating the shape of the Fermi-LAT excess in the energy range $124.7-133.4$ GeV
taken from Ref.~\cite{Su:2012ft}.\label{fig:spatial}}
\end{figure}

In the following, we leave aside the question about the origin of a possible
offset in connection with a dark matter explanation. Instead, under the
tentative hypothesis that dark matter is indeed responsible for the Fermi
feature, we discuss the impact of the apparent morphology on decaying versus
annihilating dark matter. The flux of photons arriving at the Earth per unit of
energy and solid angle from dark matter decay or annihilation is given by the
line-of-sight integral over the dark matter density or the density squared,
respectively, along the direction of observation (see e.g.~\cite{Garny:2010eg}),
\begin{equation}
\frac{dJ_\gamma}{dE d\Omega} = \frac{1}{4\pi}\delta(E-E_\gamma) \left\{ 
\begin{array}{ll}\displaystyle
\frac{1}{\tau_{\gamma\nu}m_{DM}}\ {\textstyle\int\limits_{l.o.s.}}ds\,\rho_{dm}(r) & \mbox{decay}\\[2.5ex]
\displaystyle
 \frac{2\sigma v_{\gamma\gamma}}{ m_{DM}^2}\ {\textstyle\int\limits_{l.o.s.}}ds\,\frac12\,
 \rho_{dm}(r)^2 & \mbox{annihilation}

\end{array}\right.\;,
\end{equation}
where $r=\sqrt{(r_0-s\cos\xi)^2+(s\sin\xi)^2}$, $r_0\simeq 8.5$kpc is the distance
of the Earth from the Galactic center, and $\xi$ the angle measured with respect
to the direction towards the center of the Galaxy. As a reference, we consider the
density profiles given, respectively, by a generalized Navarro-Frenk-White
(NFW)~\cite{Navarro:1995iw} or Einasto profile~\cite{Navarro:2003ew},
\begin{equation}
\rho_{dm}(r) \propto \frac{1}{(r/r_s)^\alpha(1+r/r_s)^{3-\alpha}},\ \exp\left(-\frac{2}{\alpha_E}(r/r_s)^{\alpha_E}\right)\,,
\end{equation}
with $\alpha_E=0.17$ (see e.g.\,\cite{Navarro:2008kc}), $\alpha=1$ and scale
radius $r_s=20$kpc. We also consider
two strongly peaked distributions with $\alpha=(1.3,1.7)$, that may arise from
adiabatic contraction. All profiles are normalized such that the local density
is $\rho_0=0.4\GeV/\cm^3$, except for $\alpha=1.7$ where we use $\rho_0=0.28\GeV/\cm^3$
corresponding to the maximal allowed $1\sigma$-value when taking observational
constraints from microlensing and measurements of the Galactic dynamics into
account \cite{Iocco:2011jz}. Note that for all other profiles the local density
$\rho_0=0.4\GeV/\cm^3$ lies within the $1\sigma$-allowed region.

In Ref.~\cite{Su:2012ft} it has been argued that the morphology of the Fermi
excess at $\approx 130$GeV is consistent with dark matter annihilation assuming
the Einasto or NFW profile, possibly with $\alpha$ slightly larger than one. In
order to obtain a similar emission profile for decaying dark matter, a much more
peaked dark matter distribution would be necessary. In order to illustrate this
point, we show in Fig.~\ref{fig:spatial} the expected spatial distribution for
the contracted NFW profiles. For comparison we also show the corresponding shape
for annihilating dark matter, and the count maps taken from~\cite{Su:2012ft}. 
Qualitatively, it seems that even for the strongly contracted profile with
$\alpha=1.7$ the expected distribution for decaying dark matter is too flat,
especially for the profile along the Galactic plane. One may speculate that,
in the perpendicular direction, the excess could be less peaked. However, a
definite statement is presently impossible due to statistical limitations. 

\begin{figure}
\vspace*{-1cm}
\begin{center}
 \includegraphics[width=0.85\textwidth]{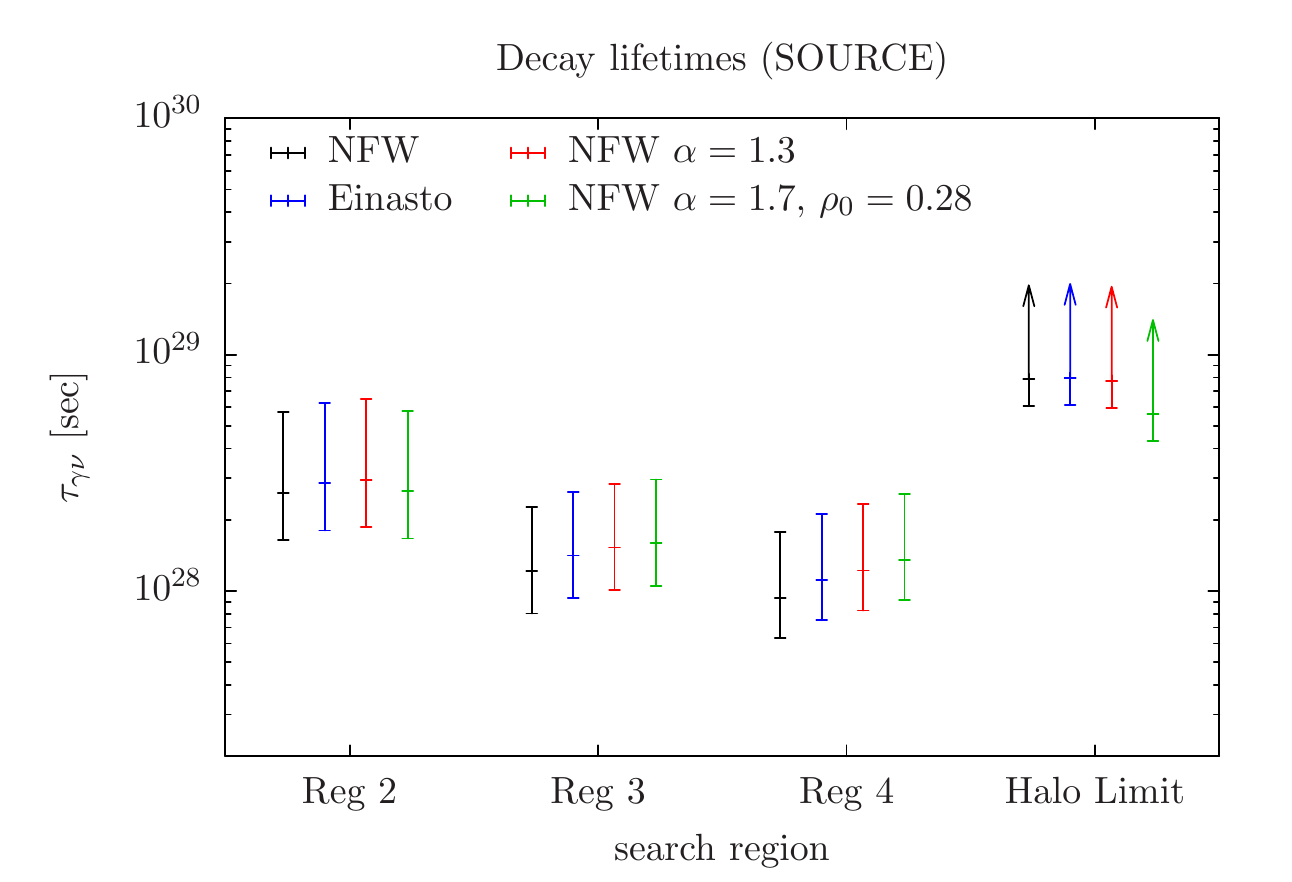}\\[1.5ex]
 \includegraphics[width=0.85\textwidth]{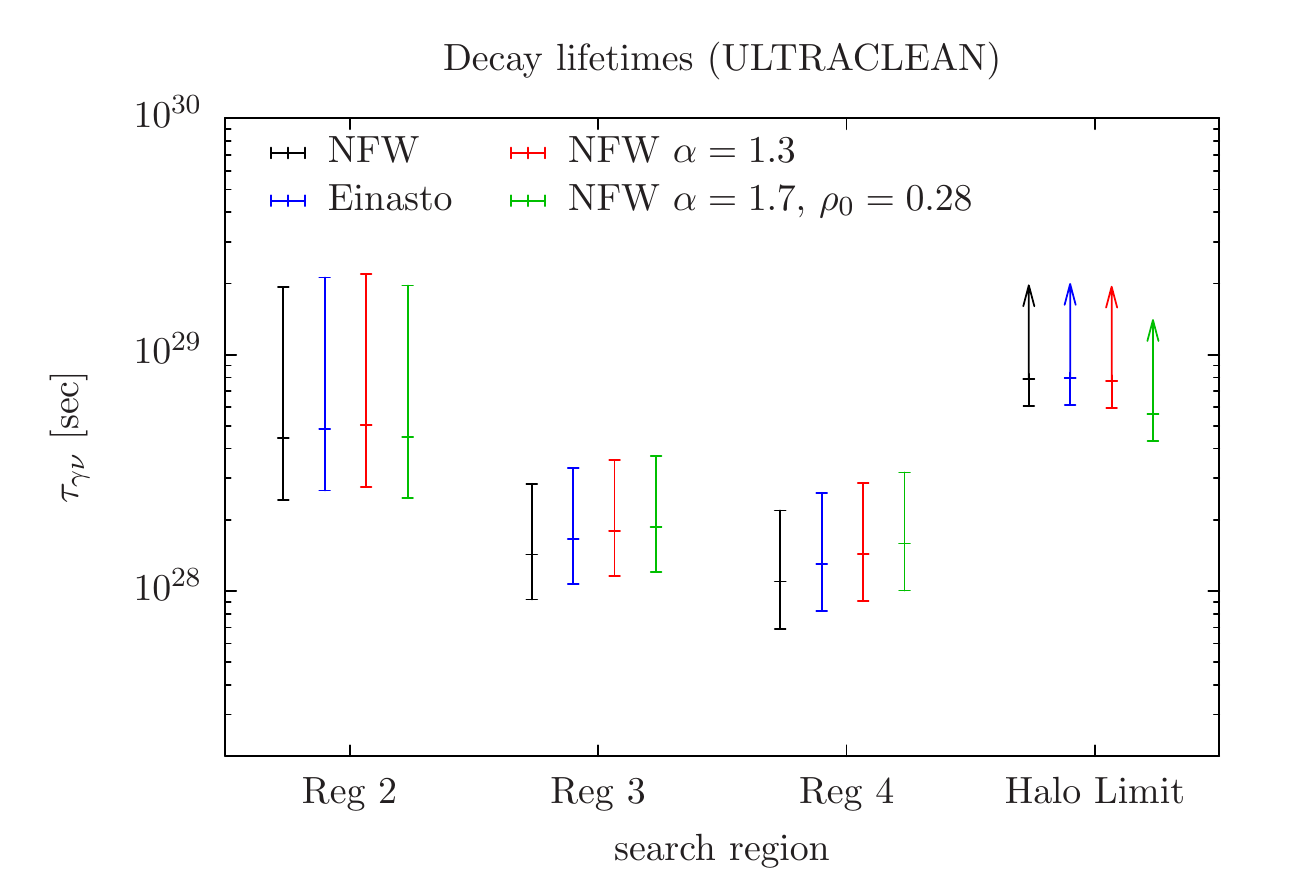}
\end{center}
 \caption{\label{fig:pbarp}\small Best fit values for the partial dark matter
lifetime $\tau_{\gamma\nu}$ for a decay into monochromatic photons, obtained
for the search regions 2, 3 and 4 defined in Ref.~\cite{Weniger:2012tx}, for
various dark matter profiles. The error bars show the 95\%C.L. regions. Also
shown are the lower limits on the partial lifetime from the most recent Fermi
LAT analysis (the lower error bars include the estimated systematic uncertainty)
\cite{Ackermann:2012qk}, that are based on a wide search region (Halo Limit). The
upper and lower figures correspond to the two Fermi data samples used
in~\cite{Weniger:2012tx}.}
\end{figure}

For decaying dark matter, one would expect also a comparably large signal arising
from the Galactic halo, where no excess has been observed. Therefore, it is
instructive to compare the halo limit with the required lifetime to explain the
excess in the central region. The partial dark matter lifetimes $\tau_{\gamma\nu}$
that are required in order to explain the feature in the Fermi gamma-ray spectrum
present within various search regions defined in~\cite{Weniger:2012tx}, and for
the various assumptions of the radial dark matter distribution, are shown in
Fig.~\ref{fig:pbarp}. Although the best-fit values for the considered search
regions are compatible with each other, they are in tension with the lower limits
on the lifetime obtained by the recent analysis of the Fermi collaboration
\cite{Ackermann:2012qk}, also shown in Fig.~\ref{fig:pbarp}. Note that the Fermi
analysis is based on a rather wide search region, and therefore sensitive to the
dark matter flux originating in the halo of the Milky Way. Due to the linear
scaling of the flux with the dark matter density in the case of decaying dark
matter, one typically expects a larger flux from the halo compared to the center
than for annihilating dark matter. Thus, for the case of decaying dark matter
the gamma-ray line should have been observed also within the search region used
by the Fermi collaboration, provided that the radial distribution of dark matter
can be well described by the considered NFW or Einasto profiles. The discrepancy
is less pronounced for the contracted profiles which are more cuspy towards the
Galactic center. On the other hand, for annihilating dark matter, the cross section
required to explain the Fermi gamma-ray feature is compatible with earlier upper
limits assuming NFW or Einasto profiles~\cite{Weniger:2012tx}.

Clearly, this tension disfavors the explanation of the excess in terms of decaying
dark matter. However, because of the large uncertainty of the actual dark matter
distribution close to the Galactic center as compared to the halo
\cite{Bertone:2004pz,Bergstrom:2012fi,Springel:2008cc}, and since one cannot
exclude large systematic uncertainties at present \cite{Ackermann:2012qk}, we
believe that it would be premature to rule out decaying dark matter as an
explanation of the Fermi excess based on this tension.

\section{Conclusion}

In light of tentative evidence for a gamma-ray line observed by Fermi-LAT
near the Galactic center, we have investigated the consequences for several
prototype scenarios of decaying and annihilating dark matter, motivated by
supersymmetric models with gravitino,  higgsino or wino-like LSP. We find
that, independently of the actual dark matter distribution, the consistency
of continuum and monochromatic contributions to the photon spectrum from dark
matter decay or annihilation requires a branching ratio into monochromatic
photons larger than $\BR_\gamma \gtrsim 0.5\%$. Both higgsino and wino dark
matter can be ruled out because of a too large continuum flux, independently of
the production mechanism, while gravitino dark matter with wino NLSP is
compatible. We have also investigated constraints arising from the primary
antiproton flux, which depend on the adopted propagation model. The resulting
limits are comparable to the ones from continuum gamma-rays for annihilating
dark matter, but can be more important for decaying dark matter. The morphology
of the tentative excess, if confirmed, could yield valuable information on the
distribution of dark matter close to the Galactic center. When assuming
conventional NFW or Einasto profile functions, the required partial lifetimes for
decaying dark matter $\tau_{\gamma\nu} \sim (1-3)\times 10^{28}$\,s are in
tension with lower limits obtained from the Galactic halo. This tension is less
pronounced for more cuspy contracted profiles. If decaying gravitino dark matter
is indeed responsible for the excess, the dark matter density should be enhanced
in the Galactic center region compared to conventional models.

\section*{Acknowledgements}

The authors thank S.~Bobrovskyi, T.~Bringmann, L.~Covi, J.~Hajer, A.~Ibarra,
A.~Morselli, A.~Ringwald and C.~Weniger for helpful discussions. We thank
M.~Grefe for useful comments and are grateful to J.~Hajer for cross-checking
the Appendix. This work has been supported by the German Science Foundation (DFG)
within the Collaborative Research Center 676 ``Particles, Strings and the Early
Universe''.

\section*{Appendix}

In the following we collect the formulae needed to evaluate the various
branching ratios of gravitino LSP decays in supersymmetric models with 
bilinear R-parity breaking.

The couplings of the gravitino field $\psi_\nu$ to matter and gauge fields 
fields are given by
\cite{Wess:1992cp,Moroi:1995fs},
\begin{equation}\label{gravcoup}
  \mathcal{L} = \frac{i}{\sqrt{2}M}
  \left(\overline{\chi} \gamma^\nu \gamma^\mu (D_\mu \phi) \psi_\nu 
        + \mathrm{c.c.}\right) -
  \frac{1}{4 M}\overline{\lambda} \gamma^\nu\sigma^{\mu\rho}
   \psi_\nu F_{\mu\rho}.
\end{equation}
Here $\chi$ stands for the left-handed lepton and higgsino doublets $l$, $h_u$,
$h_d$, and $\phi$ for the corresponding scalar leptons and Higgs fields
$\tilde{l}$, $H_u$, $H_d$;  $F_{\nu\rho}$ denotes the fields strengths of
$U(1)_Y$ and $SU(2)$ gauge interactions, $B_{\nu\rho}$ and $W^I_{\nu\rho}$,
and $\lambda$ the corresponding gauginos $b$ and $w^I$. 

In the case of bilinear R-parity breaking the transition from weak to mass
eigenstates has been worked out in detail in \cite{Bobrovskyi:2010ps,Bobrovskyi:2012xy}.
Inserting these field transformations in Eq.~(\ref{gravcoup}) and shifting
the Higgs fields around their vacuum expectation values, one finds
for the R-parity breaking gravitino couplings
\begin{align}\label{Rviol}
{\cal L}_3 \supset 
\frac{i}{2\Mp}
\{&\left(\kappa_{hi}\partial_\mu h + i\kappa_{zi} m_Z Z_\mu\right)
\bar{\nu}_i\gamma^\nu\gamma^\mu \psi_{\nu} 
+ i \sqrt{2}\kappa_{wi} m_W W^-_\mu \bar{e}_i\gamma^\nu\gamma^\mu \psi_{\nu}
\nonumber\\ 
& + i \left(\xi_{zi} \partial_\mu Z_\nu 
+ \xi_{\gamma i} \partial_\mu A_\nu\right)
\bar{\nu}_i\gamma^\lambda \sigma^{\mu\nu} \psi_{\lambda} \nonumber\\
& + i \sqrt{2} \xi_{wi} \partial_\mu W^-_\nu 
\bar{e}_i\gamma^\lambda\sigma^{\mu\nu} \psi_{\lambda} \} + \mathrm{h.c.} \ .
\end{align}
The couplings $\kappa$ and $\xi$ are determined by the R-parity breaking
parameters $\zeta_i$, the gaugino masses $M_{1,2}$ and the $\mu$ parameter
\cite{Bobrovskyi:2010ps,Bobrovskyi:2012xy},
\begin{align}\label{coefficients}
\kappa_{hi} &= \zeta_i \ ,\\
\kappa_{zi} &= \zeta_i\left(1
+\sin{2\beta}\frac{m_Z^2\left(M_1\cos^2{\theta_W} + M_2\sin^2{\theta_W}\right)}
{M_1 M_2 \mu}\right) \ ,\\
\kappa_{wi} &= \zeta_i\left(1
+\sin{2\beta}\frac{m_W^2}{M_2 \mu}\right) \ ,\\
\xi_{\gamma i} &= \zeta_i\cos{\theta_W}\sin{\theta_W}
\frac{m_Z\left(M_2 - M_1\right)}{M_1 M_2} \ ,\\
\xi_{z i} &= -\zeta_i
\frac{m_Z\left(M_2\sin^2{\theta_W} + M_1\cos^2{\theta_W}\right)}{M_1 M_2} \ ,\\
\xi_{w i} &= -\zeta_i
\frac{m_W}{M_2} \ ,
\end{align}
where $\theta_W$ is the weak angle. For a gravitino LSP one has
$M_1,M_2,\mu \gtrsim m_{3/2}$.
Note that the gauge boson couplings $\xi$ satisfy the relation
\begin{equation}
\cos{\theta_{\rm W}} \xi_{Zi} + \sin{\theta_{\rm W}} \xi_{\gamma i} = \xi_{W i} \ .
\end{equation}
The gravitino couplings (\ref{Rviol}) have previously been obtained
in an effective Lagrangian approach \cite{Buchmuller:2009xv}\footnote{Note the 
change in notation compared to \cite{Buchmuller:2009xv}. We also
corrected a typo in Eq.~(2.7) of \cite{Buchmuller:2009xv} in the prefactor
of our Eq.~(\ref{Rviol}).}. The couplings
$\xi = \mathcal{O}(m_Z/m_{3/2})$ correspond to dimension-6 operators and the 
leading terms of the couplings $\kappa$ to a dimension-5 operator, which
gives
\begin{align}
\Gamma(\psi_{3/2}\rightarrow h\nu_i) &\simeq
\Gamma(\psi_{3/2}\rightarrow Z\nu_i) \simeq
\frac{1}{2}\Gamma(\psi_{3/2}\rightarrow W^+e^-_i)\ ,
\end{align}
for $m_Z /m_{3/2} \ll 1$. The subleading contributions 
$\mathcal{O}(m_Z^2/m_{3/2}^2)$ to $\kappa$ correspond to dimension-7 operators
which were not considered in \cite{Buchmuller:2009xv}.

Based on the calculations of \cite{Covi:2008jy,Grefe:2008zz} one can now
obtain the partial two-body gravitino decay widths by matching coefficients.
The result reads
\begin{align}
\Gamma(\psi_{3/2}\rightarrow h\nu_i) &=
\frac{m_{3/2}^3}{384\pi\Mp^2}\kappa_{hi}^2\beta_h^4\ ,\label{bh}\\
\Gamma(\psi_{3/2}\rightarrow \gamma\nu_i) &=
\frac{m_{3/2}^3}{64\pi\Mp^2}|\xi_{\gamma i}|^2\ ,
\label{bgamma}\\
\Gamma(\psi_{3/2}\rightarrow Z\nu_i) &=
\frac{m_{3/2}^3}{384\pi\Mp^2}\beta_Z^2
\left(\kappa_{zi}^2 H_Z 
+ 16 \frac{m_Z}{m_{3/2}}{\rm Re}(\kappa_{zi}\xi_{z i}) G_Z 
+ 6 |\xi_{z i}|^2 F_Z\right)\ , \label{bz}\\
\Gamma(\psi_{3/2}\rightarrow W^+e^-_i) &=
\frac{m_{3/2}^3}{192\pi\Mp^2}\beta_W^2
\left(\kappa_{wi}^2 H_W 
+ 16 \frac{m_W}{m_{3/2}}{\rm Re}(\kappa_{wi}\xi_{w i})G_W 
+ 6 |\xi_{W i}|^2 F_W\right)\label{bw}\ .
\end{align}
The functions
$\beta_a$, $H_a$, $G_a$ and $F_a$ ($a=h,Z, W$) 
are given by \cite{Covi:2008jy}
\begin{align}
\beta_a &= 1 - \frac{M_a^2}{m_{3/2}^2}\ ,\\
H_a &= 1 + 10 \frac{M_a^2}{m_{3/2}^2}
+ \frac{M_a^4}{m_{3/2}^4}\ ,\\
G_a &= 1 + \frac{1}{2} \frac{M_a^2}{m_{3/2}^2}\ ,\\
F_a &= 1 + \frac{2}{3} \frac{M_a^2}{m_{3/2}^2}
+ \frac{1}{3}\frac{M_a^4}{m_{3/2}^4}\ .
\end{align}

The decay width $\Gamma(\psi_{3/2}\rightarrow \gamma\nu)$ is of particular
interest since it determines the strength of the gamma line at the end of
the continuous spectrum. Neglecting corrections $\mathcal{O}(m_Z^2/m_{3/2}^2)$,
one finds for the branching ratio
\begin{equation}\label{line}
\BR_\gamma \simeq \frac{3\pi\alpha}{2\sqrt{2}G_F}
\frac{(M_1-M_2)^2}{M_1^2M_2^2}\ ,
\end{equation}
where $\alpha$ is the electromagnetic fine-structure constant and $G_F$ is 
Fermi's constant.

\begin{figure}
\vspace*{-1cm}
\begin{center}
 \includegraphics[width=0.85\textwidth]{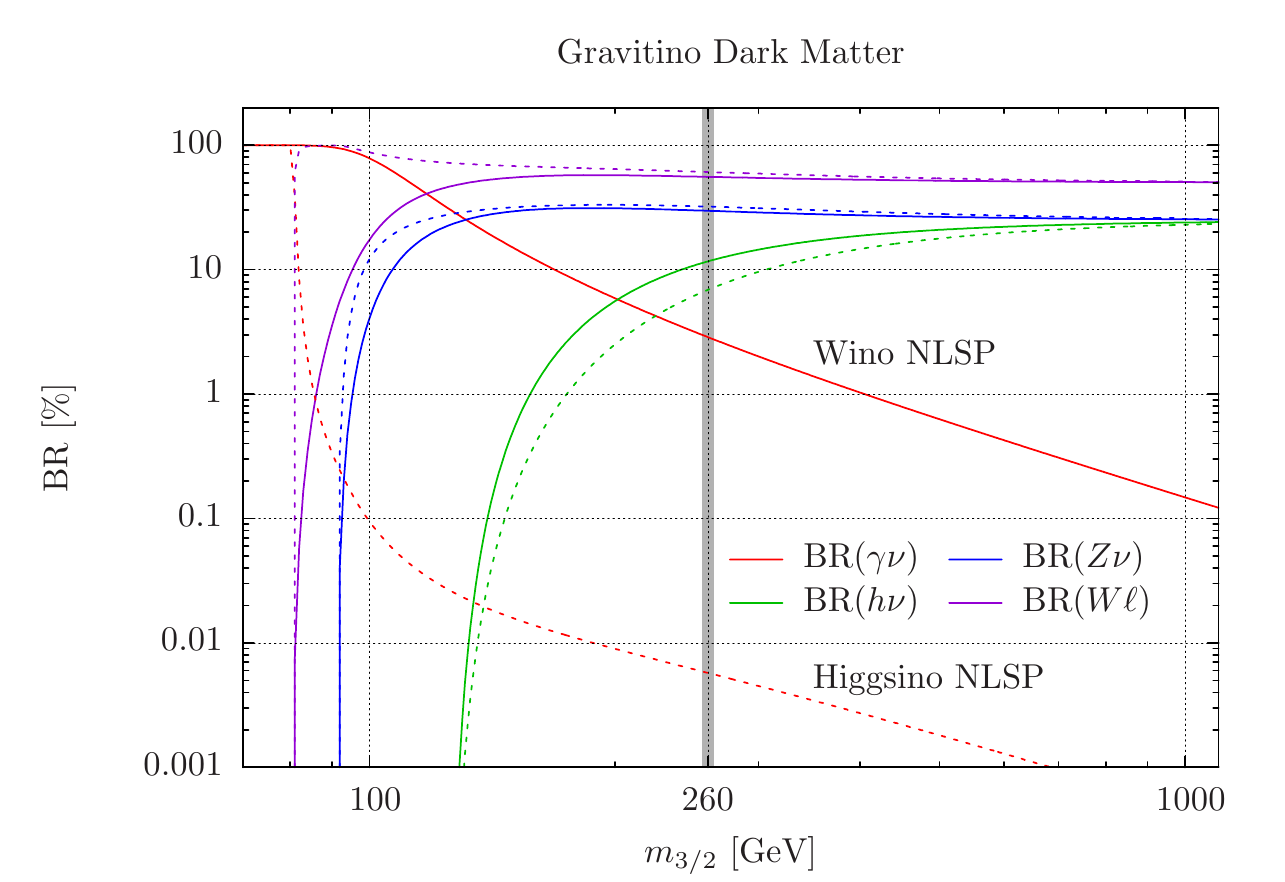}
\end{center}
 \caption{\small Branching ratios of two-body gravitino decays for two
representative examples. Wino NLSP: $M_2 = 1.1\ m_{3/2}$,
$M_1 = \mu = 10\ m_{3/2}$, and higgsino NLSP: $\mu = 1.1\ m_{3/2}$,
$M_1 = 10\ m_{3/2}$, $M_2 = 1.9\ M_1$ ($\tan\beta=10$ and $m_h=125$\,GeV in both cases).}
\label{branching}
\end{figure}

In Figure~\ref{branching} the different branching ratios are shown as functions
of the gravitino mass for the two representative cases of wino and higgsino NLSPs.


\end{document}